\documentclass{aastex63} 

\usepackage{xcolor}
\usepackage{natbib}
\usepackage{amsmath, amsfonts, amssymb}
\usepackage{threeparttablex, longtable, booktabs}
\usepackage[frak=esstix]{mathalpha}
\usepackage{appendix}
\usepackage{comment}
\usepackage{tabularx} 
\usepackage{dirtytalk}

\submitjournal{AJ}
\accepted{September 2, 2025}

\graphicspath{{./}}

\begin{document}

\title{
High-Pass Filtering and Gaussian Process Regularization: Stellar Activity Characterization Techniques Applied to the 55 Cancri Planetary System
}

\correspondingauthor{Justin Harrell}
\email{harrell@udel.edu}

\author[0000-0001-6771-4583]{Justin Harrell} 

\affiliation{University of Delaware \\
Bartol Research Institute \\
Department of Physics and Astronomy \\
Newark, DE 19716, USA}

\author[0000-0002-8796-4974]{Sarah E. Dodson-Robinson} 
\affiliation{University of Delaware \\
Bartol Research Institute \\
Department of Physics and Astronomy \\
Newark, DE 19716, USA}


\begin{abstract}

Doppler planet searches are complicated by stellar activity, through which cyclical changes in the host star's photosphere and chromosphere can mask or mimic planetary signals. A popular technique for modeling stellar activity is to apply a quasiperiodic Gaussian process (GP) kernel, which provides a flexible model with rigorous error propagation. However, observers must guard against overfitting, as a GP may be flexible enough to subsume other signals besides the one it is intended to model. To counteract overfitting, we introduce a curvature-penalizing objective function for fitting GP models to long-term magnetic activity cycles. We also demonstrate that a Gaussian filter can be an effective method of detrending radial velocities (RVs) so that shorter-period signals can be extracted even in the absence of a mathematical model of the long-term trend. We apply our methods to the heavily studied 55 Cancri system, fitting Keplerian orbits plus the GP activity-cycle model. We show that a 4-Keplerian model that includes planets b, c, e, and f combined with a GP for the activity cycle performs at least as well as the widely agreed-upon 5-planet system with its own GP activity model. Our results suggest that the existence of planet d cannot be established from the RVs alone; additional data are required for confirmation.

\end{abstract}


\section{Introduction}
\label{sec:intro}

There is a rich and complex history of radial velocity (RV) exoplanet detections around the G8 dwarf 55 Cancri A \citep[55 Cnc; $V = 5.95$, $d = 12.3~$pc, $M = 0.9 M_{\odot}$;][]{vonBraun2011}.\footnote{The binary companion star 55 Cnc B is an M dwarf with mass $M = 0.26 M_{\odot\ }$\citep{Alonso-Floriano2015}.} One of the earliest exoplanet discoveries was 55 Cnc b, a Jupiter-sized planet with a $14.6$-day period \citep{Butler1997}. Additional observations by \citet{Marcy2002} revealed another Doppler signal with period $P \sim 14$ years, which they attributed to a second gas giant orbiting at $\sim 5.5$ AU (planet d). This marked the first observation of an extrasolar Jupiter-like gas giant. \citet{Marcy2002} also discovered a $44.3$ day signal due to another Jupiter-mass planet orbiting at a much closer separation of $0.25$ AU (planet c). \citet{Dawson2010} confirmed the existence of planet e, a super-Earth at a period of $0.74$ days, correcting the initial alias detection of $P = 2.8$ days by \citet{McArthur2004}. 
The alias resulted from the 1~day$^{-1}$ spectral window artifact: $1/2.8~\text{day}^{-1} = 1/0.74~\text{day}^{-1} - 1~\text{day}^{-1}$. The confirmation of planet e made 55 Cnc the first extrasolar system known to harbor four planets. Soon after the original announcement of 55 Cnc e, independent observations of a $260$-day signal suggested the existence of planet f \citep{Wisdom2005, Fischer2008}.

With the $< 1$-day period of the innermost planet, researchers were motivated by the high geometric transit probability of $\sim25\%$ \citep{Dawson2010} to conduct an intense photometric observation campaign of 55 Cnc. The results of this successful endeavor were presented by \citet{Demory2011} and \citet{Winn2011}. At the time, 55 Cnc e was the only transiting super-Earth orbiting a star visible to the naked eye. The Rossiter–McLaughlin (RM) effect for in-transit RV measurements was initially undetected by \citet{Lopez2014} and predicted to be negligible by \citet{Bourrier2014}, as its expected RV modulation was $< 0.5$ ms$^{-1}$. An EXtreme Precision Spectrograph (EXPRES) observing campaign by \citet{Zhao2022} corroborated the earlier predictions by detecting an RM amplitude of only $0.41\substack{+0.09 \\ -0.10}$ ms$^{-1}$, a factor of 2--3 smaller than the smallest instrumental error bars from the multi-telescope planet discovery campaign \citep[complete dataset presented by][]{Bourrier2018}. Investigations into the composition and habitability of 55 Cnc e include \citet{Endl2012, Ehrenreich2012, Demory2016B, Demory2016A, Ridden-Harper2016, Esteves2017, Bourrier2018}. 

The thorough reanalysis of the 55 Cnc system by \citet{Bourrier2018} used the two decades of archival RV data from \citet{Fischer2008, Endl2012, Lopez2014, Butler2017}, as well as their own additional observations from SOPHIE. \citet{Bourrier2018} were the first to model the solar-like magnetic cycle alongside the planet orbits. As a result, the major differences between their results and the previous literature were found in the orbital parameters of the planet with the longest period. \citet{Bourrier2018} reported a planet-d period of $5574\substack{+94 \\ -89}$ days, compared with earlier values of $4909 \pm 30$ and $5285 \pm 4.5$ days by \citet{Endl2012} and \citet{Fischer2008}, respectively. However, \citet{Bourrier2018} modeled the magnetic cycle with a Keplerian orbit of amplitude $15.2 \substack{+1.6\\-1.8}$ ms$^{-1}$ and period $P = 3822.4 \substack{+76.4\\-77.4}$ days. A Keplerian orbit is a strictly periodic function, and there is significant evidence that stellar magnetic cycles are instead quasiperiodic \citep{Baliunas1995, Saar1997, Dumusque2012, GomesdaSilva2012}.

Here we extend the work of \citet{Bourrier2018} by introducing a quasiperiodic activity-cycle model. Our new analysis of 55~Cnc allows us to introduce two mathematical techniques. First, the RV oscillations with periods $\gtrsim 1000$~days are filtered out of the time series using a Gaussian high-pass filter. After iteratively estimating the orbital frequencies of planets b, c, e, and f using generalized Lomb-Scargle periodograms of the short-period residuals, we model the original, unfiltered time series using a combination of Keplerian orbits and a curvature-penalized Gaussian process for the activity cycle. Our goals are twofold: (1) to demonstrate mathematical techniques for detrending data and modeling long-term quasiperiodic signals, and (2) to characterize the 55 Cnc activity cycle and, as much as possible, parse out its RV signature from that of planet d. 





In \S \ref{sec:observations} we describe the archival RV dataset that is the basis for our work. In \S \ref{sec:detrending} we explore the process of detrending RVs: \S \ref{sec:detrending_motivation} outlines the frequency domain-based motivations for detrending, 
while \S \ref{sec:GF-detrending} describes Gaussian filter (GF) detrending and explains its advantages. In \S \ref{sec:GP_modeling}, we present our curvature-penalized implementation of a quasiperiodic Gaussian process (GP), with particular emphasis on magnetic activity cycles. We discuss our approach to Keplerian fitting in \S \ref{sec:fitting}, while in \S \ref{sec:model_comparison} we compare the merits of a 5-Keplerian model and a 4-Keplerian model that lacks planet d. We 
provide our conclusions in \S \ref{sec:conclusions}.

\section{Observations}
\label{sec:observations}

We use the aggregated RV dataset presented by \citet{Bourrier2018} (see Figure \ref{fig:GF_RVdetrending}), which combines data from HARPS \citep{Pepe2003}, HARPS-N \citep{Cosentino2012}, Keck HIRES \citep{Vogt1994}, Hamilton Spectrograph at Lick Observatory \citep{Vogt1987}, the HRS and Tull spectrographs at McDonald Observatory \citep{Bramall2012, Tull1995}, and the SOPHIE spectrograph at Haute-Provence Observatory \citep{Perruchot2008}. \citet{Bourrier2018} binned the data over a timescale of 30 minutes, which helps to average out the shorter-period stellar signals such as oscillations and granulation \citep{Dumusque2011}. The binned RV time series has $N = 782$ observations spanning over $9251$ days with a median interval between observations of $\Delta t = 1.031$ days. 



\section{Methods: High-pass Filtering}
\label{sec:detrending}

The first step toward detecting and characterizing the signals from planets b, c, e, and f is to remove the long-term trend caused by the combination of magnetic activity and planet d from the data. In this section we explain why trend removal is necessary and demonstrate detrending with a Gaussian filter, which does not require a mathematical model of the long-term variability.

\subsection{Motivations to Detrend}
\label{sec:detrending_motivation}

 
Typical convective stars have a long-term activity cycle in which changes in the magnetic field strength and polarity cause modulation in the number of star spots and plages. Since these regions of intense magnetic flux suppress convective blueshift, the magnetic activity cycle produces a quasiperiodic Doppler shift \citep{Gray2009, Meunier2010}. There is considerable evidence that stellar activity can masquerade as a planetary signal \citep{Desidera2004, Carolo2014, Saar1997, Hatzes2002, Desort2007, Boisse2011, Robertson2014_A, Robertson2014_B, Newton2016, SuarezMascareno2017, Rajpaul2021}. Previous work suggests the 55 Cnc magnetic cycle should have RV perturbations on the order of $\sim 10$ ms$^{-1}$ \citep{GomesdaSilva2012, Meunier2013}. See \citet{Dumusque2012} for a detailed review of the sources of stellar activity. 

The long-term magnetic cycle can dominate the Lomb-Scargle periodogram, making signal identification at higher frequencies very difficult \citep[for examples of how long-period signals can conceal shorter-period ones, see][]{ Hinshaw2003, Campante2011, Haley2014, Bugnet2022}. This phenomenon is known as spectral leakage: a fraction of the power at each frequency will not stay confined to the correct frequency bin but will ``leak" across the frequency domain, thus reducing the accuracy of the overall power spectrum estimate \citep[Section~2.1]{SDR_Haley2024}. Even linear trends and large mean values can contribute power to the zeroth frequency bin and produce crippling leakage, which is why most authors will subtract off the mean of a time series before performing Fourier analysis \citep[e.g.][]{shumwaystoffer}.

\subsection{High-pass Filtering}
\label{sec:GF-detrending}

Choosing an appropriate mathematical model for the activity cycle can be challenging, as our knowledge of the physics underlying magnetic dynamos is still evolving \citep{Fischer2016, Dumusque2017, Cegla2019, Hatzes2019}. The type of detrending we perform here is to separate the long-period and short-period signals into two separate time series, which are called trend and residual. (In the 55~Cnc dataset, we roughly define ``long-period'' as $P \gtrsim 1000$ days and ``short-period'' as $P \in [1, \sim 100]$ days.) The observer can search for planets in the residual time series, then add back in the trend before fitting a complete time-domain model that includes the activity cycle. Since the Fourier transform is a linear operator, the trend and residual time series together preserve the frequency-domain representation of the original time series. Detrending is especially useful when the time series records a signal with $f < 2 \mathcal{R}$, where $\mathcal{R}$ is the Rayleigh resolution limit:
\begin{equation}
    \mathcal{R} = \frac{1}{t_{N-1} - t_0}.
\label{eq:Rayleigh_resolution}
\end{equation}
Such signals are unresolved from zero frequency \citep[e.g.]{Godin1972, ramirezdelgado25}, so they reduce the periodogram's dynamic range while providing limited insight into oscillatory phenomena.

We use a low-pass Gaussian filter (GF) to calculate the trend, which is a smoothed version of the original time series, then subtract the trend from the data to get the residual. Our trend estimate via the GF at time $t=t_n$, $\mathrm{RV} = y_{\text{GF}}[t_n]$ (where $n$ is the time index), is computed via an adaptation of convolution to unevenly spaced timestamps:
\begin{equation}
\begin{aligned}
    g_n &= \exp{\Big\{-\frac{(t-t_n)^2}{2 \sigma^2}\Big\} } \\
    y_{\text{GF}}[t_n] &= \frac{\sum_{n = 0}^{N-1} g_n y_n}{\sum_{n = 0}^{N-1} g_n} .
\label{eq:Gauss_filter}
\end{aligned}
\end{equation}

\begin{figure}
    \centering
    \includegraphics[width=1.0\textwidth]{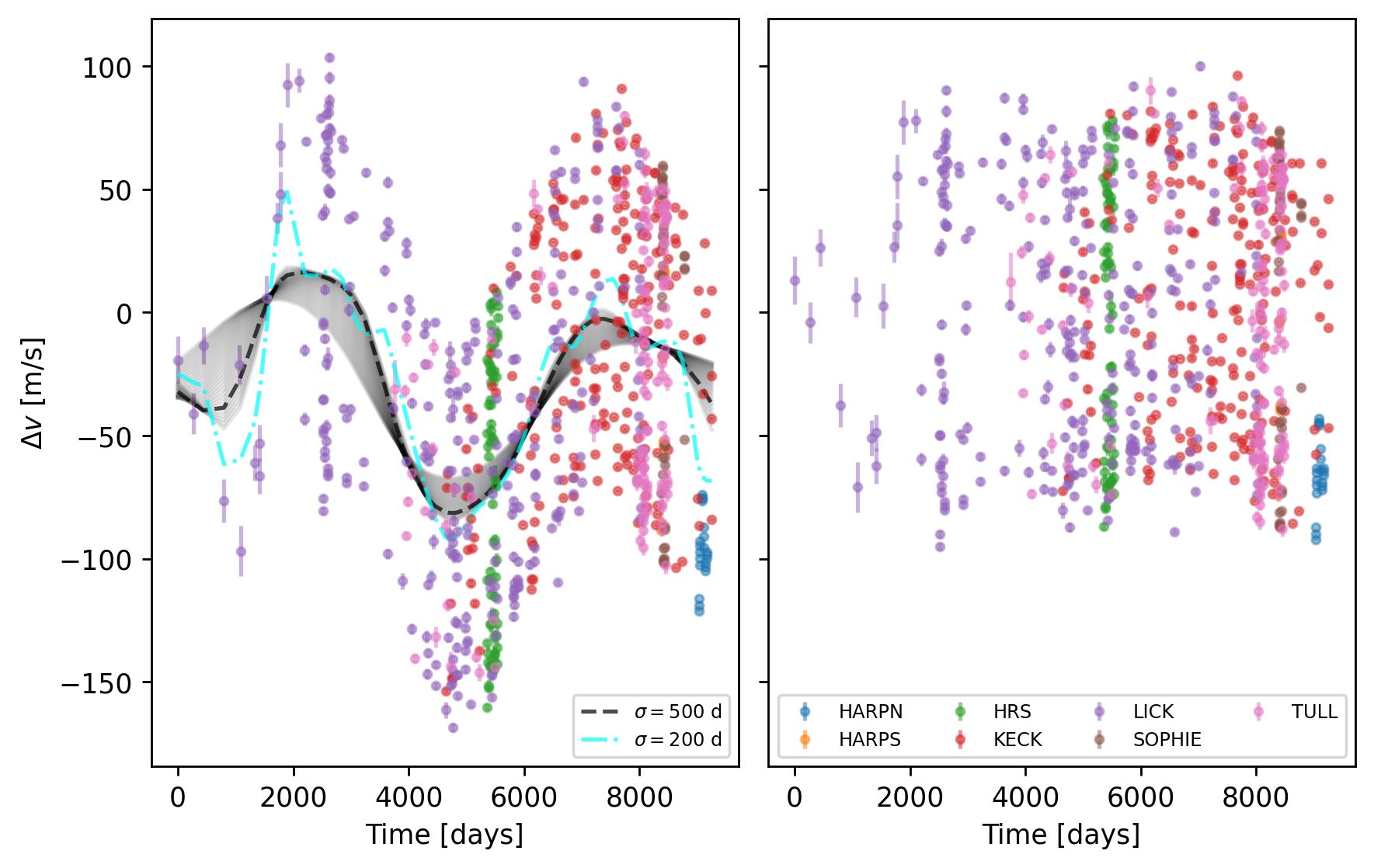}
    \caption{{\bf Left:} Long-term trend (dashed black line) computed using a Gaussian filter (GF) via Equation \ref{eq:Gauss_filter} superposed onto the 55 Cnc RV time series. The Gaussian filter used here has a width of $\sigma = 500$ days. The gray shaded region is comprised of 30 GF trends for a reasonable range of $\sigma \in [400, 1000]$ days (sampled uniformly). We include a GF trend with a poorly chosen width of $\sigma = 200$ days (cyan dash-dotted line) for comparison. The GF long-term trend represents the weighted moving average of RVs across this high-pass filter. {\bf Right:} The detrended RV time series after the long-term trend is subtracted out.}
    \label{fig:GF_RVdetrending}
\end{figure}

Using Equation \ref{eq:Gauss_filter} we compute the GF-estimated long-term trend (dashed black line) displayed in Figure \ref{fig:GF_RVdetrending}. Here we utilize a GF of fixed width $\sigma = 500$ days,\footnote{A fixed Gaussian filter width may be inappropriate for datasets with large gaps in observation. In this case, a segmented averaging \\ system may be preferable; an example can be seen in \citet{SDR2022}.} which we chose because it produces a smooth long-term trend without extracting noticeable short-term fluctuations. We find that $y_{\text{GF}}[t_n]$ changes only marginally for a filter width range of $\sigma \in [400, 1000]$ days, giving us confidence that our long-term trend is not acutely sensitive to the choice of filter width. The gray shaded region is comprised of 30 GF trends with uniformly sampled filter widths $\sigma \in [400, 1000]$ days. We also provide an example of a GF trend computed with a poor choice of $\sigma = 200$ days (cyan dash-dotted line), which exhibits undesirable short-term fluctuations that risk filtering out planetary signals. While it is clear from visual inspection that the long-term trend timescale is $\gtrsim 3000$~days, $\sigma = 200$ days produces a trend in which the time elapsed between the global maximum and the subsequent local maximum is only $730$ days. As a general heuristic, we recommend choosing $\sigma$ such that the GF removes variation on timescales of order $1/2\mathcal{R}$, in accordance with the Rayleigh resolution limit. Using this heuristic therefore excludes a GF with $\sigma = 200$ days.

\begin{figure}
    \centering
    \includegraphics[width=0.9\textwidth]{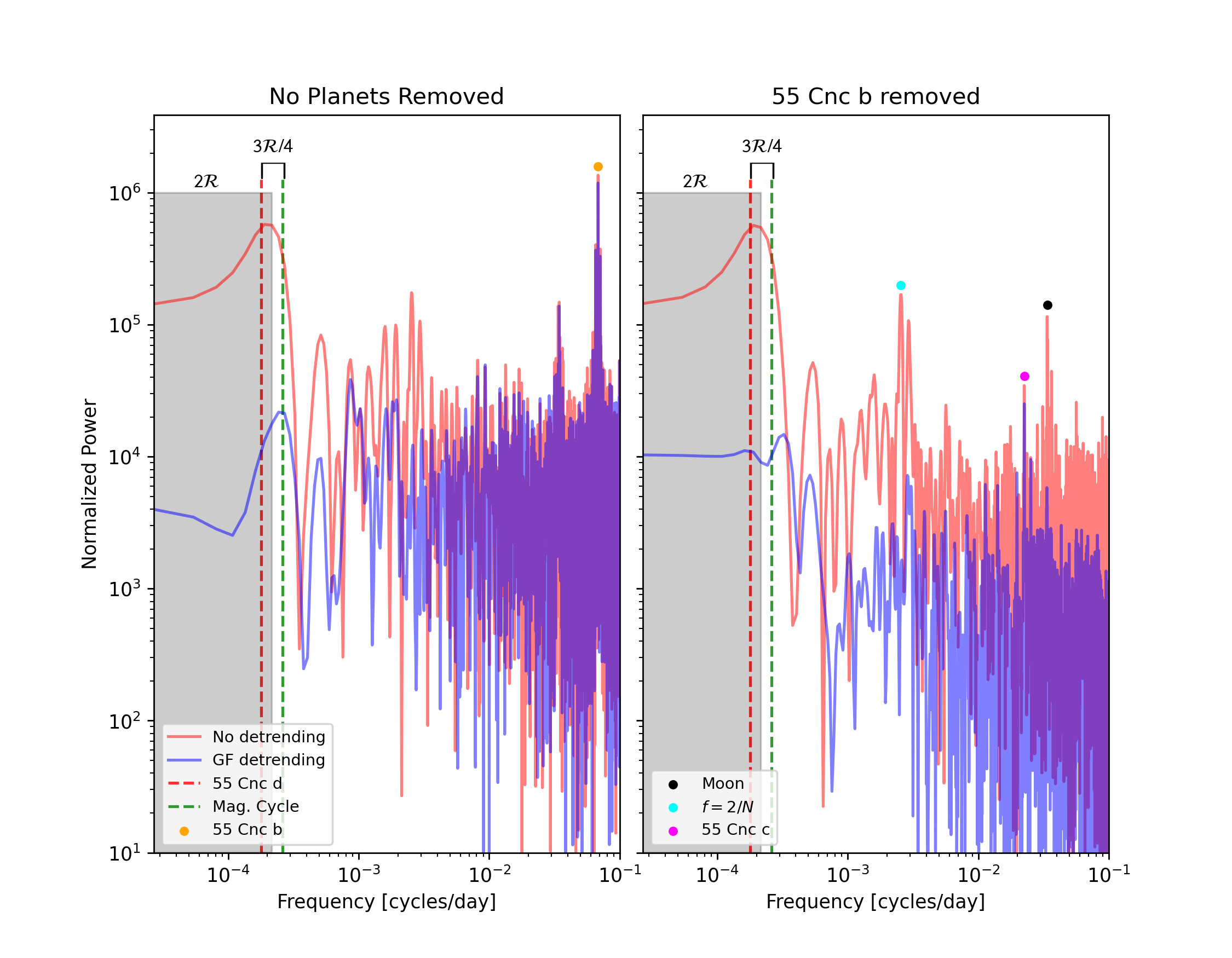}
    \caption{{\bf Left:} Periodograms of the undetrended (red) and detrended (blue) RVs, before removing planet b (left) and after removing planet b (right). The shaded gray rectangle has width $2 \mathcal{R}$ and shows the interval in which signals are not distinguishable from zero frequency. The black bracket has width $3\mathcal{R}/4$, which is the frequency separation of planet d's orbit and the activity period given by \citet{Bourrier2018}.}
    \label{fig:detrending_spectra}
\end{figure}

In Figure \ref{fig:detrending_spectra}, we display the GLSP of the RVs before and after the GF detrending given by Equation \ref{eq:Gauss_filter}.\footnote{All periodograms presented in this paper are normalized using Parseval's theorem so that their average value is equal to the time series variance.} The periodograms in the left panel of Figure \ref{fig:detrending_spectra} show the effect of Gaussian filter detrending on the RVs depicted in Figure \ref{fig:GF_RVdetrending}. The GLSP computed from the GF-detrended RVs (blue) has reduced power by a factor of $\sim 25$ for $f \lesssim 10^{-4}$ d$^{-1}$ from the GLSP of the un-detrended RVs (red). But even before detrending, planet b produces the strongest signal in the periodogram, which we mark with a orange dot.

The importance of detrending is most evident after we subtract out the time-domain model of planet b (we discuss Keplerian fitting in \S \ref{sec:fitting}). The right-hand panel of Figure \ref{fig:detrending_spectra} compares periodograms of residuals after removing planet b (red) with residuals after first removing planet b, then detrending with the GF (blue). Without detrending, a significant portion of power lies within $0 < f < 2\mathcal{R}$ (gray shading), shrouding other signals by reducing dynamic range. The red curve also has alias peaks corresponding to the moon's synodic orbital frequency (black) and $f = 2 / N$ (light blue). However, the highest peak in the blue curve is at the frequency of planet c (pink), which is meaningful for an observer who is searching for more planets. After periodogram peaks corresponding to planets have been identified, the observer may use the original RVs to fit a global model that includes Keplerians plus a physics-based model of stellar activity.



\section{Methods: Stellar Activity Modeling with a Gaussian Process}
\label{sec:GP_modeling}

There is considerable evidence that stellar magnetic cycles are quasiperiodic \citep{Saar1997, Dumusque2012, GomesdaSilva2012, Robertson2013, Fuhrmeister2023}. \citet{Baliunas1995} analyzed $111$ main sequence stars and found that the average level of chromospheric activity and rotational variability both change on evolutionary timescales, with Sun-like stars like 55 Cnc exhibiting occasional periods of reduced magnetic flux in Maunder minimum \citep{wright04}. The Sun's Schwabe cycle is another example: although the $\sim 11$-year periodicity appears in a wide range of observables such as sunspot number, total solar irradiance, and 10.7cm radio flux, the cycles vary in amplitude, time between solar minimum and maximum, and duration \citep{hathaway15}. Observations of the Sun also show asymmetries in Sun spot evolution between northern and southern hemispheres \citep{Chowdhury2019, El-Borie2020}.

Gaussian processes (GPs) are commonly used to model quasiperiodic variation due to rotation or stellar activity \citep[e.g.][]{Haywood2014, Foreman-Mackey_2017, Angus18}. The general idea behind using GPs is to provide a flexible and tractable mathematical framework for Bayesian inference modeling of phenomena that are not fully deterministic. GPs are particularly useful for modeling stochastic processes with rigorous error propagation, even with limited knowledge of the underlying process \citep{Rajpaul2015}. 

\subsection{The Quasiperiodic Kernel}
\label{sec:GP_kernel}

Using the \texttt{Python} package \texttt{George}\footnote{\url{https://github.com/dfm/george}} \citep{george_tech_paper}, we construct a quasiperiodic GP kernel as informed by Equation 27 from \citet{Rajpaul2015}:
\begin{equation}
    k_{\text{QP}}(t_i, t_j) = K^2 \text{exp}\Bigg\{ - \frac{ \sin^2{[\pi(t_i - t_j)/P]}}{2 \lambda_p^2} - \frac{(t_i - t_j)^2}{2 \lambda_e^2}  \Bigg\} , 
\label{eq:GP_ker}
\end{equation}
where $K$ is the amplitude, $P$ is the period of oscillation, $\lambda_p$ is the dimensionless length scale of periodic variation (often called a correlation scale), and $\lambda_e$ is an evolutionary time scale in the units of $t$. This particular kernel has both a well-studied history of effectively modeling quasiperiodic stellar RV modulation \citep[e.g.][]{farr18, bortle21, cabot21} and a physical motivation for its mathematical form \citep{aigrain12, Rajpaul2015}. The first term in Equation \ref{eq:GP_ker} captures periodic variation through the non-stationary exponential-sine-squared kernel and the second term describes amplitude modulation through the stationary squared exponential kernel. See \citet{Rasmussen_GP_textbook} for a detailed description of GPs.

The major difficulty in implementing a GP to describe a magnetic activity cycle comes from the requirement that the model be smooth on dynamo timescales \citep[$>1$-year;][]{Baliunas1995, Robertson2013, hathaway15}. Optimized models found by minimizing the GP's negative log-likelihood function (NLL) may end up fully explaining the RV variance, thereby producing short-term fluctuations in addition to the long-term variation. For recent examples of GP overfitting, see \citet{Blunt2023, Beard2024}. One may counteract this tendency to overfit by using priors to heavily restrict the GP parameter space, but this strategy does not necessarily yield a desirable optimization, as one or more of the best-fit parameters may fall on an imposed boundary. This was a common occurrence in our exploratory modeling of 55 Cnc, and poses a challenging question: How can one be confident that the upper and lower bounds on GP parameters are physically motivated? If the optimization algorithm fits GP parameters to values near the bounds, it is possible that the best-fit parameters are only optimal in the restricted parameter space and do not accurately reflect the underlying magnetic cycle of the star.

We note that \citet{suarezmascareno23}, who also discuss the overfitting problem, use  unbounded lognormal prior distributions on amplitudes and jitter for their stochastically excited, damped harmonic oscillator kernel \citep{Foreman-Mackey_2017} in order to ensure a smooth activity model. But as \citet{Rasmussen_GP_textbook} and \citet{curvature_penalty_textbook} point out, model optimization solely through the NLL-prior combination has some pitfalls. Without additional restrictions on the model, its $\chi^2$ per degree of freedom can be reduced to 0 by letting the model interpolate the data. The optimal value of reduced $\chi^2$ is unity; a value near zero indicates overfitting \citep[e.g.][]{bevington03}.

\subsection{The Curvature-Penalizing Objective Function}
\label{sec:objective_function}


Our approach to avoiding GP overfitting while allowing for a broad parameter space is to include a curvature penalty in our objective function (often called regularization). See \citet{curvature_penalty_textbook} for a general description of regularization or \citet[Section~6.2]{Rasmussen_GP_textbook} for its specific use with GPs. 
The curvature penalty is proportional to the sum of the squared second derivative of the model prediction and is added to the NLL.

Let $y_n$ denote the observed RV at time $t_n$ with reported uncertainty $\sigma_{y, n}$. Let $\Delta y_n$ denote the residual at $t_n$ after subtracting all Keplerian orbits. That is, for a model comprised of an integer number of Keplerians $S$, each predicting an RV of $\mathcal{K}_{s, n}$ at time $t_n$, let 
\begin{equation}
    \Delta y_n \equiv y_n - \sum_{s = 0}^{S-1} \mathcal{K}_{s, n} . 
\label{eq:keplerian_residual}
\end{equation}
We minimize a curvature-penalized $\chi^2$ objective function, $\mathcal{L}(\boldsymbol{\theta})$, onto $\Delta y_n$:
\begin{equation}
    \mathcal{L}(\boldsymbol{\theta}) = -\frac{1}{2} \sum^{N-1}_{n = 0}  \frac{(\Delta y_n - \tau_n)^2}{\sigma_{y, n}^2}  - \frac{1}{2} \alpha \sum^{M}_{m = 1} \Bigg( \frac{\tau_{m+1} - 2\tau_m + \tau_{m-1}}{\Delta t_{\text{GP}}^2} \Bigg)^2 ,
\label{eq:CW_NLL}
\end{equation}
where $\tau_m$ is the stellar activity-induced Doppler shift predicted by $k_{\text{QP}}$ (Equation \ref{eq:GP_ker}) at time $t_m$, computed as the mean of the conditional predictive distribution.

Importantly, the transformation from covariance into the time domain need not be done on the grid of RV observation times. We therefore take advantage of the central difference Taylor approximation to the second derivative of a function sampled on a uniformly spaced grid. In principle, the second derivative could be estimated at the same timestamps as the time series $t_n$, resulting in $M = N-2$. However, we found no advantages to this sampling cadence for GPs that are smooth on long timescales. In order to speed up the computation of $\mathcal{L}(\boldsymbol{\theta})$ 
during model optimization, we compute the GP predictions on a more sparse grid $t_m$ with uniform sampling $\Delta t_{\text{GP}}$, with $M = N/3$. This yields a GP sampling cadence of $\Delta t_{\text{GP}} = 35.58$ days, which is more than two orders of magnitude shorter than the previously estimated activity cycle period \citep{Bourrier2018}.

The parameter $\alpha$ dictates the degree to which GP curvature is penalized, and can be tuned to the desired tradeoff between minimizing prediction error and suppressing short-term variability. From a Bayesian perspective, the curvature penalty in Equation \ref{eq:CW_NLL} is equivalent to using the standard NLL while placing a prior proportional to $\exp{[-\frac{\alpha}{2}\int (\frac{d^2\tau}{dt^2})^2 ]} $ over the space of all smooth functions. Of course, the challenge is choosing an appropriate $\alpha$ for the given model. The aforementioned literature on regularization exclusively treats uniformly sampled data, so their recommendations about selecting $\alpha$ do not necessarily translate to irregularly sampled time series.

As a general heuristic, we identified a useful range of values for $\alpha$ in which the minimum produces a GP prediction that is more variable than one would expect from a long-term magnetic activity cycle \citep[ i.e.\ similar to the cyan curve in Figure \ref{fig:GF_RVdetrending}; see also][]{Baliunas1995}. Conversely, the maximum penalty yields a GP prediction that is insufficiently variable to the point of producing a final RV residual that indicates the model has not fully explained the data. There exists a reasonable range of curvature penalties between these two extremes that will yield similar and appropriate fits, and we recommend users narrow the range of $\alpha$ values they consider by visualizing the GP prediction. One may also select $\alpha$ via an algorithmic approach such as cross-validation \citep[Section~3.2]{curvature_penalty_textbook}.

As we later discuss in \S \ref{sec:model_comparison}, we construct both a 4-Keplerian and a 5-Keplerian model, each with a GP to model stellar activity. We use $\alpha = 2\cdot10^8 \ \text{day}^4\text{m}^{-2}\text{s}^2$ for the 4-Keplerian model and $\alpha = 3\cdot10^9 \ \text{day}^4\text{m}^{-2}\text{s}^2$ for the 5-Keplerian model. We find these to be the minimum curvature penalties required to prevent short-term interpolating of each GP, similar to the short-term fluctuations exhibited by the GF trend of $\sigma=200$ days in Figure \ref{fig:GF_RVdetrending}. The large order of magnitude results from the fact that $\alpha$ scales with $\Delta t_{\text{GP}}^4$ (recall $\Delta t_{\text{GP}} = 35.58$ days). These values are also consistent with economics research on the Hodrick-Prescott filter, which uses a similar curvature penalizing objective function \citep{HP1997}. Reviews such as \citet{Morten_adjusting_HP_filter} recommend using $\alpha = 1600 \ \Delta t$, which gives $\alpha \approx 2.5 \cdot 10^9 \ \text{day}^4\text{m}^{-2}\text{s}^2$ for the GP sampling cadence described above. The 5-Keplerian + GP model has a higher value of $\alpha$ than its 4-Keplerian counterpart because the planetary component has more flexibility, which tends to drive curvature in the GP. In general, the best value of $\alpha$ increases with model complexity.

\section{Methods: Keplerian Fitting}
\label{sec:fitting}

In addition to the GP model of stellar activity, our time-domain model includes Keplerian orbits of all planet candidates that are identified in the frequency domain (\S \ref{sec:freq_domain}). 
To detect planets and estimate their periods before parameter optimization with the Nelder-Mead method, we use the following procedure:
\begin{description}
    \item[(1) Fit for instrumentation effects] Before including any Keplerians in our time-domain model, we fit for the zero-point offsets of each instrument. The RV contributions from Keplerian orbits and zero-point offsets are calculated with the \texttt{kepmodel}\footnote{\url{https://obswww.unige.ch/~delisle/kepmodel/doc/index.html}} package. 
    \item[(2) Detrend RVs] After fitting and subtracting the zero-point offsets, we use a GF to detrend as explained in \S \ref{sec:GF-detrending}.
    \item[(3) Fit Keplerian] If the global maximum of the GLSP computed from the residual RVs reaches a statistically significant level (statistical significance is further explained in \S \ref{sec:freq_domain}), we add a Keplerian to our model. We define a Keplerian orbit using the period $P$, RV semi-amplitude $K$, eccentricity $e$, argument of periastron $\omega$, and mean anomaly $L_0$ at the first observation timestamp $t_0$. 
    \item[(4) Compute residual and repeat step (3)] We find the residual RVs by subtracting each Keplerian orbit, then compute the GLSP. We repeat step (3) until no signals are statistically significant. 
\end{description}
Steps 1 and 4 use the publicly available fitting tools of the Data \& Analysis Center for Exoplanets (DACE). After all planets have been detected and removed from the GF-detrended RVs, we add the GF trend back in to model the magnetic cycle as explained in \S \ref{sec:GP_modeling}. We use Nelder-Mead nonlinear optimization with adaptive parameters \citep{Gao2012} to find the best-fit Keplerian orbits and GP parameters. Initial guesses for planet periods are obtained through frequency-domain analysis, which we describe in the next section. Instrumental jitter is estimated from the residuals after model fitting; see Appendix \ref{sec:appendix} for details.




\subsection{Frequency Domain Analysis}
\label{sec:freq_domain}

Here we describe how planet candidates are identified in Step 3 of the procedure outlined above. After using a GF of width $\sigma = 500$ days to detrend the 55 Cnc RVs (Step 2, Figure \ref{fig:GF_RVdetrending}), 
we turn to the frequency domain to identify planetary signals. We have two criteria for considering a signal within the GLSP as significant. First, the signal must have a false-alarm probability (FAP) $< 1\%$. We estimate the FAP $p$ using the $1-p$ percentile of the distribution of peak power from $10^6$ bootstrap iterations. 
Second, for a model consisting of multiple Keplerian orbits, we must also assess whether adding another planet results in over-fitting. To do so, we compute the Bayesian Information Criterion (BIC) \citep{Schwarz1978}: 
\begin{equation}
    \text{BIC} = \gamma \ln{(N)} - 2\ln{(\hat{L}_{\text{max}})},
\label{eq:BIC}
\end{equation}
where $\gamma$ is the number of parameters in the model, $N$ is the number of observations in the time series, and $\hat{L}_{\text{max}}$ is the model's estimated maximum likelihood, as computed by \texttt{kepmodel}. A more complex model is only allowed if the gain in explained variance outweighs the penalty for increasing the number of free parameters. If the $\text{BIC}$ increases after adding a Keplerian or GP, this step likely represents overfitting and is rejected from the model \citep{Schwarz1978}. For each Keplerian fit in Step 3, we compute the change in BIC ($\Delta\text{BIC}$) due to the additional Keplerian. The Keplerian is only accepted into the model if $\Delta\text{BIC} \geq 10$, since \citet{Delta_BIC_10} found that this gives a high degree of confidence that the additional Keplerian improves the model. 

\begin{figure}
    \centering
    \includegraphics[width=0.9\textwidth]{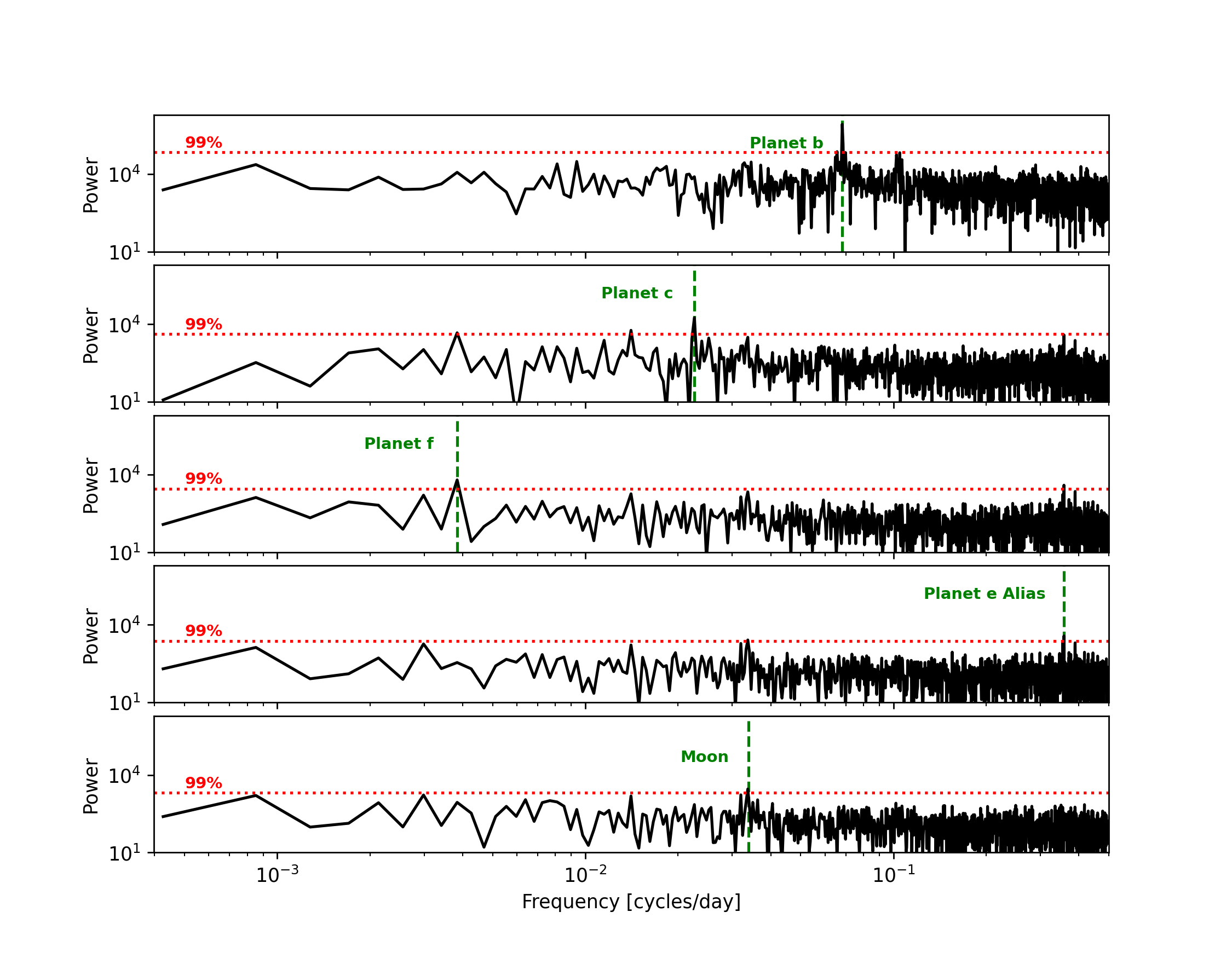}
    \caption{Periodograms for each residual RV during the Keplerian fitting process. At this stage, our fitting is done on the GF-detrended RVs. In order of removal, we fit Keplerians to planets b, c, f, and e. The horizontal dotted line represents the 1\% false-alarm probability, as estimated by a $10^6$ iteration bootstrap simulation of the periodogram.}
    \label{fig:GLS_4p_RV_GFdetrended}
\end{figure}

Figure \ref{fig:GLS_4p_RV_GFdetrended} shows the GLSPs after fitting the following model components: instrumental offsets (top), planet b followed by detrending with the Gaussian filter
(second from top), planet c (middle), planet e (second from bottom), and planet f (bottom).
At each model fitting step, all parameters are optimized simultaneously.
The periodogram peaks from planets b, c, f, and the alias of e all cross the bootstrapped 99\% significance threshold. After those planets' removal from the detrended RV time series, the only significant peak in the GLSP of the residuals is at the lunar synodic frequency. This merely reflects the observing cadence associated with ``bright-time'' RV observation campaigns, and indicates that there are no other signals to account for in the GF-detrended RVs. 

For the binned RV time series, the planet e frequency of $f_e = 0.736550 \, \text{d}^{-1}$ lies above the approximate Nyquist frequency $\hat{f_{N}} \approx 1 / 2 \bar{\Delta t} = 0.48 \,\text{d}^{-1}$. For this reason, we only display the planet e alias $f_{\text{alias}} = f_e - 1 \: \operatorname{day}^{-1} = 0.357 \,\text{d}^{-1}$ in Figure \ref{fig:GLS_4p_RV_GFdetrended}. An alias is a spurious signal that arises in a power spectrum estimate due to its convolution with the spectral window (see \S \ref{sec:detrending_motivation}). Aliases are therefore solely the byproduct of sampling cadence; see \citet{McClellan1999} for a thorough description of aliasing. The planet e alias is the strongest signal after removing planet c, and disappears after we subtract a Keplerian with the true orbital frequency of planet e, which we take from \citet{zhao23}. Validation of planet e is not of particular concern here since it has transit confirmation and an observed Rossiter-McLaughlin effect \citep{Demory2011, Winn2011, zhao23}. Aliases can be detected through analyzing the spectral window of the time series, and 55~Cnc reminds us of the necessity of this step \citep{Dawson2010}.


The periodograms displayed in Figure \ref{fig:GLS_4p_RV_GFdetrended} unsurprisingly lack evidence for planet d, since they were computed on the GF-detrended RVs. The proposed planet d with $P = 5574\substack{+94 \\ -89}$ days \citep{Bourrier2018} lies within the frequency range affected by the GF detrending. It also lies within the zero-frequency bin of the GLSP, as its separation from $f = 0$ is less than two Rayleigh resolution units $\mathcal{R}$, which is the minimum frequency separation required to statistically distinguish two signals in a periodogram \citep{Godin1972, SDR2022, ramirezdelgado25}.
Since planet d is statistically indistinguishable from both zero frequency and the activity cycle, we cannot characterize it in the frequency domain. 
Instead, we first fit the orbits for planets b, c, e, and f, then add the GF long-term trend $y_{\text{GF}}$ (see Equation \ref{eq:Gauss_filter})
to the residual time series. This sequence of operations isolates the long-period oscillations so that they can be fitted with an activity model and, optionally, planet d (see \S \ref{sec:model_comparison}). 




\subsection{Disentangling planet d and the activity cycle}
\label{section:d_activity}

\citet{Bourrier2018} modeled the combination of planet d and the activity cycle using two Keplerians with periods $P_d = 5574.2^{+93.8}_{-88.6}$ days and $P_{\rm mag} = 3822.4^{+76.4}_{-77.4}$ days. However, solar and stellar activity cycles are not strictly periodic: amplitude variability and phase drift are evident in all observables, including RV, S-index, H$\alpha$, photometry, and more \citep[e.g.][]{Baliunas1995, Saar1997, robertson13, suarezmascareno16, fuhrmeister23}. 
To capture amplitude variability, any model of an activity cycle built from strictly periodic functions {\it must include at least two components}---in other words, two Keplerians are required to fit the long-term RV variability of 55~Cnc even without planet d.

Let us then advance the audacious hypothesis that planet d is a spurious artifact of fitting strictly periodic functions to a quasiperiodic phenomenon. 
To test whether the RV data support the existence of planet d, we will compare two time-domain models. The first consists of planets b, c, e, and f plus a GP model of the activity cycle, while the second has all of the above plus planet d. We discuss the merits of the competing models in \S \ref{sec:model_comparison}.

\section{RV Results: Four-planet vs.\ five-planet model comparison}
\label{sec:model_comparison}

We now proceed with a head-to-head comparison between two different models of the RV time series. The first RV model, which we call 4pGP, has planets b, c, e, and f but lacks planet d, and the quasiperiodic GP of Equation \ref{eq:GP_ker} describes the magnetic activity cycle. The second model, called 5pGP,  includes all five reported planets plus the GP for stellar activity. Model 5pGP is our counterpart to the work of \citet{Bourrier2018}, where the only difference (beyond our use of regularization) is that we describe the activity cycle with a quasiperiodic GP rather than a strictly periodic Keplerian.

\subsection{Best-fit models}
\label{subsec:bestfit}

\begin{figure}[h]
    \centering
    \includegraphics[width=0.9\textwidth]{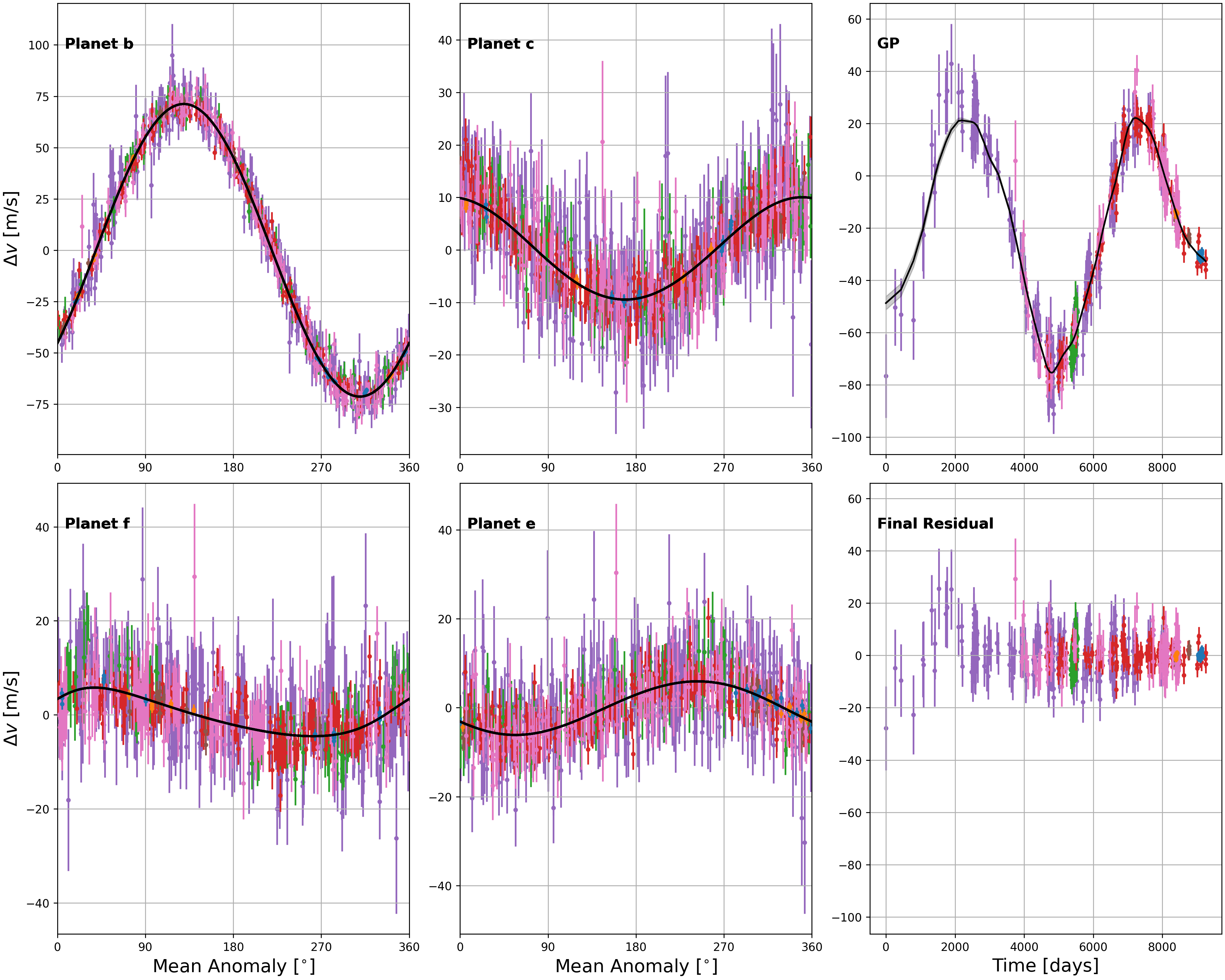}
    \caption{Best-fit Keplerian orbits for planets b, c, f, and e. In the bottom right panel, we display the final RV residual after fitting these four planets in addition to the magnetic cycle. Note that the rightmost panels have time rather than orbital phase on the horizontal axis and matching y-axis ranges. Other panels do not necessarily use the same y-axis range. Error bars depicted here are the quadrature sum of errors reported by \citet{Bourrier2018} and the instrument-dependent jitter we find in Appendix \ref{sec:appendix}. This plot uses the same color scheme as Figure \ref{fig:GF_RVdetrending}.}
    \label{fig:res_panel_4pGP}
\end{figure}

Our results show that the exact mathematical description of the star's long-term variability barely affects the characterization of planets b, c, and e, and has a modest affect on the predicted orbit of planet f. In Table \ref{table:best_fits}, which lists the best-fit Keplerian orbital parameters and their uncertainties for both models,\footnote{See Table \ref{table:instrument_fits} for best-fit instrument offsets and instrumental jitter estimates.} we see excellent agreement between 4pGP and 5pGP on $P$, $K$, and $e$ for planets b, c, e, and f. (Note that the non-matching values of $\omega$ for planets with nearly circular orbits have limited physical meaning, as the argument of periastron is undefined for a circular orbit.)
Our b, c, e, and f orbits are also generally consistent with \citet{Bourrier2018}, with $P$ and $K$ agreeing within uncertainties for all planets and $e$ agreeing within uncertainties for planets b, c, and e. Only for planet f do our models record an eccentricity that is higher than that of \citet{Bourrier2018} by an amount outside the published uncertainties. 
The left-most four subplots of Figure \ref{fig:res_panel_4pGP} show the best-fit phase-folded Keplerian orbits for planets b, c, e, and f from the 4pGP model. 

Recall that our 4pGP model is a mathematical test of the plausibility of excluding planet d and instead using only a GP to explain the long-term RV variability. The top-right panel of Figure \ref{fig:res_panel_4pGP}, which depicts the residual RVs from model 4pGP with the GF trend estimate added back in, shows quasi-periodic variation consistent with an activity cycle \citep[e.g.][]{lubin21}. The black line shows the median GP prediction, which has period $4980 \pm 930$~days and amplitude $K = 12.2 \pm 6.7$~m~s$^{-1}$ (see Table \ref{table:GP_fit} for best-fit GP parameters). Our 4pGP activity cycle amplitude agrees within uncertainties with that of \citet{Bourrier2018}, who found $K = 15.2 \substack{+1.6 \\ -1.8}$~m~s$^{-1}$ for their Keplerian activity model. There are two factors that contribute to a larger uncertainty on the best-fit GP amplitude than on the semi-amplitude of the Keplerian activity model presented by \citet{Bourrier2018}. First, our uncertainty estimates come from the Nelder-Mead simplex, as opposed to the Markov Chain Monte Carlo (MCMC) estimation performed by \citet{Bourrier2018} (see \citet[Appendix~A]{NM_1965} for simplex methodology). Second, the Keplerian semi-amplitude is strictly defined and constant for all time, while the amplitude of a quasiperiodic process varies from cycle to cycle. When decorrelation is added to the activity model, a range of $K$ values can adequately reproduce same variability pattern.

\begin{figure}
    \centering
    \includegraphics[width=0.95\textwidth]{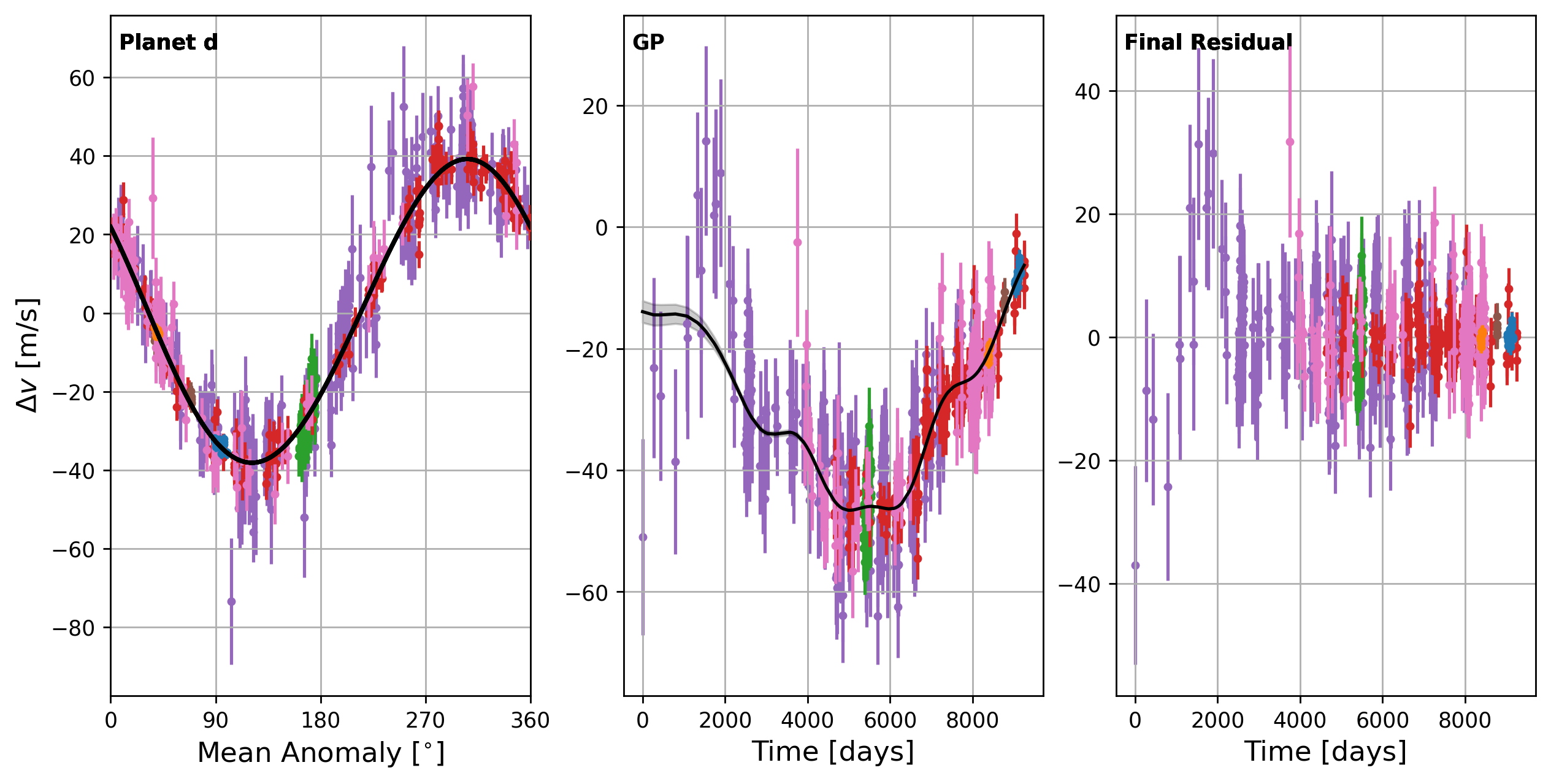}
    \caption{Features of the 5pGP model. {\bf Left:} Phase-folded best-fit orbit of planet d. {\bf Center:} Median GP prediction; note the dominance of signal drift over periodicity. {\bf Right:} residual of the 5pGP model.
    Note the changing vertical axis scales. This plot uses the same color scheme as Figure \ref{fig:GF_RVdetrending} and Figure \ref{fig:res_panel_4pGP}.}
    \label{fig:res_panel_5pGP}
\end{figure}

The most notable deviation from previous work is the orbit of planet d in our 5pGP model: we find a best-fit period of $P_d = 4600 \pm 130$~days and eccentricity $e = 0.0245 \pm 0.063$, compared with $P_d = 5574.2 \substack{+93.8 \\ -88.6}$~days and $e = 0.13 \substack{+0.02 \\ -0.02}$ from \citet{Bourrier2018}. We discuss how this change impacts potential astrometric observations of planet d in \S \ref{sec:conclusions}. In Figure \ref{fig:res_panel_5pGP} we show the 5pGP model's phase-folded RV curve of planet d (left), median GP model prediction (center), and residuals after subtracting off all five planet orbits plus the median GP prediction (right).

 
The most important conclusion from our experiments is that our 4pGP model, which treats the long-term RV variability with a single GP instead of two Keplerians, performs at least as well as the \citet{Bourrier2018} model in statistical goodness-of-fit metrics, as we will see in the next section.


\begin{table}
\centering
\footnotesize
\caption{\textbf{Best-fit Parameters for Instrumentation Effects.}}
\begin{tabular}{||c | c c c c c c c||} 
 \hline 
 \textbf{4pGP} & HARPN\textsuperscript{1} & HARPS\textsuperscript{1} & HRS\textsuperscript{2} & KECK\textsuperscript{2} & LICK\textsuperscript{2} & SOPHIE\textsuperscript{2} & TULL\textsuperscript{2} \\ [0.5ex] 
 \hline\hline
 RV Offset [ms$^{-1}$] & $27477 \pm 12$ & $27491.7 \pm 4.3$ & $28420.9 \pm 6.7$ & $-16.7 \pm 2.2$ & $27.0 \pm 3.5$ & $27461.6 \pm 6.7$ & $-22547.4 \pm 2.1$ \\ 
Jitter [ms$^{-1}$] & $0.98$ & $0.77$ & $4.4$ & $2.8$ & $6.3$ & $1.7$ & $3.7$ \\
\hline
\textbf{5pGP} & & &  & & & \\ [0.5ex] 
\hline
RV Offset [ms$^{-1}$] & $27489.7 \pm 7.6$ & $27503.9 \pm 4.7$ & $28432.4 \pm 4.5$ & $-4.7 \pm 5.0$ & $37.9 \pm 2.5$ & $27473.8 \pm 2.1$ & $-22535.4 \pm 5.3$ \\
Jitter [ms$^{-1}$] & $1.0$ & $0.78$ & $4.2$ & $2.8$ & $6.4$ & $1.7$ & $3.7$ \\
\hline\hline
\end{tabular}
\\
\bigskip
Best-fit values for instrumentation offset and jitter, found via Nelder-Mead minimization of Equation \ref{eq:CW_NLL}.

\textsuperscript{1} - Jitter approximated using Equation \ref{eq:mean_square_uncertainty} since the instrument has fewer than 25 data points.

\textsuperscript{2} - Jitter estimated using CDF fitting (see Figure \ref{fig:norm_res_distributions}) and Equation \ref{eq:time_dep_jitter}.
\label{table:instrument_fits}
\end{table}

\begin{table}
\centering
\footnotesize
\caption{\textbf{Best-fit Parameters for the 55 Cnc Planetary System.}}
\begin{tabular}{||c|c c c c c||} 
 \hline
 \textbf{4pGP} & 55 Cnc b & 55 Cnc c & 55 Cnc e & 55 Cnc f & 55 Cnc d \\ [0.5ex] 
 \hline\hline
 $P$ [days] & $14.651535 \pm 6\cdot10^{-6}$ & $44.4027 \pm 0.0045$ & $0.7365484 \pm 3.9\cdot10^{-6}$ & $259.80 \pm 0.53$ & - \\ 
 $K$ [ms$^{-1}$] & $71.2 \pm 1.6$ & $9.74 \pm 0.46$ & $6.03 \pm 0.72$ & $5.17 \pm 0.93$ & - \\
 $e$ & $0.0009 \pm 0.0049$ & $0.03 \pm 0.10$ & $ 0.03 \pm 0.20$ & $0.227 \pm 0.044$ & - \\
 $\omega$ [$^{\circ}$] & $162 \pm 11$ & $6.5 \pm 4.5$ & $2.1 \pm 1.9$ & $5.28 \pm 0.35$ & - \\
 $L_0$ [$^{\circ}$] & $5.7015 \pm 0.0064$ & $0.17 \pm 0.28$ & $1.97 \pm 0.58$ & $5.04 \pm 0.48$ & - \\ 
 $M\sin{i}$ [$M_{\text{Jup}}$] & $0.802\pm0.020$ & $0.1588\pm0.0077$ & $0.0250\pm0.0030$ & $0.148\pm0.027$ & - \\ 
 $a$ [AU] & $0.117182280\pm6.3\cdot10^{-8}$ & $0.245403\pm3.3\cdot10^{-5}$ & $0.01596151\pm1.1\cdot10^{-7}$ & $0.7968\pm0.0022$ & - \\ [1ex]
 \hline
\textbf{5pGP} & & & & & \\ [0.5ex] 
 \hline
 $P$ [days] & $14.65152 \pm 5\cdot10^{-5}$ & $44.4032 \pm 0.0018$ & $0.7365482 \pm 1.7\cdot10^{-6}$ & $259.73 \pm 0.30$ & $4600 \pm 130$ \\  
 $K$ [ms$^{-1}$] & $71.22 \pm 0.20$ & $9.79 \pm 0.23$ & $6.0 \pm 0.38$ & $5.06 \pm 0.66$ & $38.6 \pm 3.5$ \\
 $e$ & $0.0013 \pm 0.0054$ & $0.036 \pm 0.025$ & $0.035 \pm 0.024$ & $0.23 \pm 0.073$ & $0.025 \pm 0.063$ \\
 $\omega$ [$^{\circ}$] & $4.9 \pm 6.8$ & $6.49 \pm 1.27$ & $1.9 \pm 3.1$ & $5.36 \pm 0.34$ & $7.3 \pm 3.7$ \\
 $L_0$ [$^{\circ}$] & $5.698 \pm 0.024$ & $0.19 \pm 0.041$ & $1.96 \pm 0.096$ & $5.03 \pm 0.31$ & $ 2.77\pm0.11$ \\ 
 $M\sin{i}$ [$M_{\text{Jup}}$] & $0.8022 \pm0.0091$ & $0.1594\pm0.0041$ & $0.0249\pm0.0016$ & $0.145\pm0.019$ & $2.95\pm0.27$ \\ 
 $a$ [AU] & $0.11718222\pm5.5\cdot10^{-7}$ & $0.245405\pm1.3\cdot10^{-5}$ & $0.015961510\pm5.0\cdot10^{-8}$ & $0.7967\pm0.0012$ & $5.41\pm0.20$ \\ [1ex]
 \hline\hline
\end{tabular}
 \\
\bigskip
Best-fit values for $P$, $K$, $e$, $\omega$, and $L_0$ were found via Nelder-Mead minimization of Equation \ref{eq:CW_NLL}. We estimated uncertainties using the methodology given by \citet[Appendix~A]{NM_1965}, with accompanying errata. We derive $M\sin{i}$ from \citet{Lovis_Fischer2010} Equation 14 with propagated uncertainties and $a$ from Kepler's third law.
\label{table:best_fits}
\end{table}



\subsection{Mathematical models and physical origins of the long-period RV variability}
\label{subsec:resultsmeaning}

Here we evaluate the quality of our models from both a physical and a statistical perspective. We provide the best-fit GP hyperparameters for each model in Table \ref{table:GP_fit}, along with goodness-of-fit metrics. 
The bottom-right panel of Figure \ref{fig:res_panel_4pGP} shows the 4pGP residual after removing all four planets plus the mean GP prediction. The y-axis scale is identical to the above panel showing the long-period signal. Compare the 4pGP residuals to Figure 5 of \citet{Bourrier2018}: the two sets of residuals are extremely similar. 

The behavior of the 5pGP model suggests that the evidence for two distinct long-term periodicities in the RVs---planet d and the activity cycle---is not strong. The best-fit GP in the 5-planet model has $\lambda_e < P$, which means the periodic character of the magnetic activity model is suppressed compared to the aperiodic drift. However, the GP periods in the 4pGP and 5pGP models agree within uncertainties. 
Additional data, either from a longer RV time series or astrometry, are required to confirm the existence of planet d. A preliminary analysis of Hubble Space Telescope Fine Guidance Sensor data by \citet{McArthur2004} suggested an astrometric perturbation consistent with a planet d inclination of $53^{\circ}$.

\begin{table}
\centering
\footnotesize
\caption{\textbf{GP Hyperparameters and Goodness of Fit Metrics.}}
\begin{tabular}{||c | c c ||} 
 \hline 
 Quantity & 4pGP & 5pGP \\ [0.5ex] 
 \hline\hline
$P$ [days] & $4980 \pm 930$ & $5050 \pm 1720$  \\ 
$K$ [ms$^{-1}$] & $12.2 \pm 6.7$ & $4.3 \pm 3.0$  \\ 
$\lambda_e$ [days] & $5015 \pm 3950$ & $3010 \pm 1210$  \\ 
$\lambda_p$ & $0.24 \pm 0.062$ & $0.665 \pm 0.099$ \\ 
$\mathcal{L}(\boldsymbol{\theta})$ & $-6123.19$ & $-6370.33$ \\
$\alpha$ [day$^4$m$^{-2}$s$^2$] & $2\cdot10^8$ & $3\cdot10^9$ \\
Res. RMS [ms$^{-1}$] & $5.49$ & $5.82$ \\
\hline\hline
\end{tabular}
 \\
\bigskip
\textbf{Note}: The root-mean-squared (RMS) value is computed on the final RV residual, with all Keplerians, mean GP predictions, and instrument offsets subtracted from the original data.  
\label{table:GP_fit}
\end{table}

Next we compare the goodness of fit of models 4pGP and 5pGP, which have comparable $\mathcal{L}(\boldsymbol{\theta})$ of $-6123$ and $-6370$, respectively. We obtain similar final RV residuals to that of \citet{Bourrier2018} after removing all Keplerians plus the mean GP predictions. Model 4pGP has a residual RMS of $5.49$~ms$^{-1}$, while 5pGP has a residual RMS of $5.82$~ms$^{-1}$. These RMS values correspond to $1.29\bar{\sigma_J}$ for 4pGP and $1.36\bar{\sigma_J}$ for 5pGP, where $\bar{\sigma_J}$ is the mean jitter across all instruments (see Appendix \ref{sec:appendix} for details of our jitter estimation procedure).

The NLL from \citet{Bourrier2018} cannot be directly compared with the curvature-penalized $\mathcal{L}(\boldsymbol{\theta})$ used in this work. In order to compare the residual RMS and reduced~$\chi^2 = [\sum_{n=0}^{N-1} (\Delta y_n - \tau_n) / \sigma_{y,n}] / (N-\gamma)$ from 4pGP, 5pGP, and \citet{Bourrier2018}, we recreate the latter using the online DACE tools (see \citet[Table~A.2]{Bourrier2018} for model parameters). 
We find the \citet{Bourrier2018} final RV residual has an RMS of $5.38$ ms$^{-1}$ 
and reduced $\chi^2 = 0.81$.\footnote{\citet[Table~3]{Bourrier2018} indicate that their RV residual has a weighted standard deviation across all instruments of $4.33$ ms$^{-1}$, though details about the weighting are not given. We therefore make no comparisons to this metric.} Our 4pGP model has 
reduced $\chi^2 = 0.75$, while 5pGP yields 
reduced $\chi^2 = 0.80$.

We further perform model validation using the Anderson-Darling goodness-of-fit (A-D GOF) test \citep{Anderson_Darling1952, Anderson_Darling1954}, which tests the null hypothesis that a given sample is drawn from a population that follows a specified distribution. We perform this test on the residual RVs from the 4pGP and 5pGP models, using a normal distribution of unspecified mean and variance. Without considering jitter (see Appendix \ref{sec:appendix}), we find a test statistic of $3.67$ for the 4pGP model and $5.05$ for the 5pGP model. We therefore reject the null hypothesis for both models at a $99\%$ confidence level, indicating the residual RVs in models 4pGP and 5pGP are not Gaussian distributed if one does not account for jitter. With the inclusion of our jitter estimates from Table \ref{table:instrument_fits}, we obtain test statistics of $0.62$ and $1.12$ for models 4pGP and 5pGP, respectively. The test statistic of $0.62$ for model 4pGP only indicates non-Gaussian residuals at a $85\%$ confidence level, which we deem to be insufficient evidence. Interestingly, the test statistic of $1.12$ for model 5pGP indicates non-Gaussian residuals at a $99\%$ confidence level, even with the inclusion of jitter. Seeing as the jitter estimates in model 5pGP are nearly identical to those in model 4pGP, the non-Gaussian residuals likely reflect a worse overall fit in the 5-planet model as opposed to the 4-planet model---a finding echoed (though less pronounced) by reduced $\chi^2$ and RMS values.

The two-sample A-D GOF test provides another model validation technique \citep{Scholz1987}. It tests the null hypothesis that two sample distributions come from the same population, without needing to specify the distribution function of that population. For both models 4pGP and 5pGP, we compute the 2-sample A-D GOF test using the cumulative distribution functions of the best-fit model and the RV data\footnote{We approximate the null distribution using $6\cdot10^4$ random permutations.}. For both model 4pGP and 5pGP, we find a test statistic of $-1.20$ with a p-value of $0.99998$, indicating no evidence to reject the null hypothesis. The two-sample A-D GOF test therefore strongly indicate that the predictions of model 4pGP and 5pGP come from the same populations as the RV data, suggesting that both models are statistically reasonable.

By residual RMS alone, the \citet{Bourrier2018} model slightly outperforms both our 4pGP and 5pGP models. Both 4pGP and 5pGP have lower reduced $\chi^2$ values than the \citet{Bourrier2018} model, with the caveat that $\chi^2$ depends on instrumental jitter, which we estimate differently than \citet{Bourrier2018}. Considering all three goodness-of-fit metrics (RMS, reduced $\chi^2$, and likelihood), we find 4pGP to be statistically superior to 5pGP, and both models to be on par with the \citet{Bourrier2018} model. Values of reduced $\chi^2 < 1$ suggest that either jitter is overestimated or all models are slightly overfitting, which argues in favor of the model with the fewest free parameters---4pGP. The 4pGP model is additionally favored by the 2-sample A-D GOF test finding its residual to be Gaussian distributed, whereas the the 5pGP residual is not Gaussian. The conclusion that a model without planet d can explain the RVs at least as well as models that include it compels further investigation into both the activity cycle and the astrometric signal from 55~Cnc; the latter is forthcoming with Gaia \citep{kervella19, Perryman2014}.

\section{Conclusions}
\label{sec:conclusions}

We used the 55~Cnc observations presented by \citet{Bourrier2018} to demonstrate two mathematical tools---Gaussian filtering and curvature-penalized maximum likelihood---that may be widely useful in planet searches. After removing the dominant RV signal caused by planet b, we detrended the residual RVs using the GF, then used Lomb-Scargle periodograms to find planet c, the alias of planet e, and planet f at $\text{FAP} < 1\%$.
However, planet d and the activity cycle are not formally detectable in periodograms as their frequencies are 
statistically indistinguishable from each other and from zero \citep[e.g.][]{Godin1972, ramirezdelgado25}.
Without clear frequency-domain evidence for planet d, we put forth the bold hypothesis that the long-term RV variation can be solely explained by the magnetic activity cycle.

To test our hypothesis, we constructed two models: 4pGP, which includes planets b, c, e, and f plus an activity cycle described by a curvature-penalized quasiperiodic GP, and 5pGP, which includes all of the above plus an additional Keplerian for planet d. The best-fit orbital parameters of planets b, c, e, and f are minimally affected by the mathematical description of the long-period signal, and model 4pGP performs at least as well as 5pGP and the \citet{Bourrier2018} model in terms of reduced~$\chi^2$. The RVs alone do not conclusively demonstrate the existence of planet d; other observations are required to confirm its existence.

Our major deviations from previous results are in the period and amplitude of the activity cycle and the period of planet d. \citet{Bourrier2018}, who were the first to model activity alongside the five reported planets, found $K = 15.2 \substack{+1.6\\-1.8}$~m~s$^{-1}$ and $P = 3822.4 \substack{+76.4\\-77.4}$~days for the activity cycle. However, it is unclear how well a strictly periodic Keplerian orbit can describe a quasiperiodic magnetic cycle. Our 4pGP and 5pGP models both find an activity-cycle period $\sim5000$~days, but both models poorly constrain the periodicity with uncertainties of $930$ days for the 4-planet model and $1720$ days for the 5-planet model. Despite our large uncertainties, the activity cycle period of our 4pGP model is not consistent with that of \citet{Bourrier2018}. Although the activity cycle period of our 5-planet model is \textit{technically} consistent within uncertainties with that of \citet{Bourrier2018}, its decorrelation timescale of $3010\pm1210$ days---less than the period---makes such a comparison challenging. See again the center panel of Figure \ref{fig:res_panel_5pGP} for the dissimilarity between this activity model and a Keplerian. Model 4pGP has a best-fit activity cycle amplitude that is consistent with \citet{Bourrier2018} ($K = 12.2 \substack{+3.2 \\ -2.3}$~m~s$^{-1}$), while the best-fit activity cycle amplitude in model 5pGP is only $4.3 \pm 3.0$~m~s$^{-1}$. 

It is challenging to decide which new model---4pGP or 5pGP---better explains the data. 
Both models produce comparable values for reduced $\chi^2$, residual RMS, and maximum log-likelihood, and they are largely in agreement on the Keplerian parameters for planets b, c, e, and f. The 4-planet model is attractive in that it requires 5 fewer parameters to obtain slightly better goodness-of-fit metrics than the 5-planet model. 
However, our principle concern with the 5-planet model is the inability to find evidence for planet d in the frequency domain. Although model 5pGP yields a sensible final RV residual similar to prior work, a researcher following our methodology without prior knowledge of the system would have no reason to introduce a fifth Keplerian. 



We predict the deciding piece of evidence on the existence for planet d will be astrometric validation. \citet{Bourrier2018} indicate planet d has an $M\sin{i} = 3.12 \pm 0.1~M_{\text{Jup}}$ with orbital separation of $a = 5.957 \substack{+0.074 \\ -0.071}$ AU, which produces measurable astrometric deviations of $\sim1.6$ mas. Our 5pGP model estimates 55 Cnc d to have $M\sin{i} = 2.95 \pm 0.27~M_{\text{Jup}}$ and $a = 5.41 \pm 0.2$ AU, which yields a slightly smaller astrometric amplitude of $1.4$ mas. A $1.4$ mas wobble from a $V = 5.95$ magnitude star is comfortably within Gaia's astrometric accuracy of $\sim 10~\mu \text{as}$ for stars with $V \sim 7-12$ \citep{Perryman2014}. Using data from Gaia, one could compare the measured astrometry to that predicted by planet d, as the deviations should be almost entirely due to the gravitational perturbations of planet d. The only complication arises from the need to observe 55 Cnc over a time period on the order of the orbital period of 55 Cnc d \citep{Perryman2014}. Just as \citet{Marcy2002} found, this likely requires astrometric observations over a 7--10 year period to separate proper motion from the orbital influence of planet d. 


\acknowledgments
Our research was partially funded through the Bartol Research Institute of the University of Delaware. This publication makes use of The Data \& Analysis Center for Exoplanets (DACE), which is a facility based at the University of Geneva (CH) dedicated to extrasolar planets data visualisation, exchange and analysis. DACE is a platform of the Swiss National Centre of Competence in Research (NCCR) PlanetS, federating the Swiss expertise in Exoplanet research. The DACE platform is available at \url{https://dace.unige.ch}. This research also utilizes the VizieR catalogue access tool, CDS, Strasbourg, France \citep{10.26093/cds/vizier}. The original description of the VizieR service was published in \citet{vizier_origin_paper}.


\software{\\Astropy \citep{astropy:2013, astropy:2018, astropy:2022}, \\
SciPy \citep{SciPy-NMeth2020}, \\
George \citep{george_tech_paper}, \\
kepmodel (\url{https://obswww.unige.ch/~delisle/kepmodel/doc/index.html})
\\}

The \texttt{Jupyter} notebook to execute the analysis in this paper is hosted at \url{https://github.com/jaharrell/55Cnc_exoplanets} and is preserved on Zenodo at \url{https://doi.org/10.5281/zenodo.14571274} \citep{harrell_2024_55Cnc_code}. A \texttt{Jupyter} notebook explaining the general RV fitting process can be found at \url{https://dace.unige.ch/tutorials/}.


\appendix
\section{Appendix: Estimating Jitter}
\label{sec:appendix}



An important step in characterizing the RV variability is to estimate the combination of instrumental jitter beyond the photon noise contribution \citep[e.g.][]{delisle18} and stellar jitter, as the reported RV uncertainties $\sigma_{y, n}$ do not account for either.
Since the NLL of a model comprised of only Keplerians can be computed by inverting a covariance matrix,
the independent instrumental and stellar jitter can be included as free parameters in the model by adding white-noise terms $\sigma_{J,n}^2$ to the diagonal of the covariance matrix. However, using the objective function in Equation \ref{eq:CW_NLL} prevents us from fitting for jitter by replacing the denominator in the first term on the right-hand side with $\sqrt{\sigma_{y,n}^2 + \sigma_{J,n}^2}$, as there is a strong algorithmic incentive to increase the $\sigma_{J,n}$ as much as the priors allow. 
We therefore propose a post-hoc method to estimate jitter \textit{after} minimization of $-\mathcal{L}(\theta)$ as defined by Equation \ref{eq:CW_NLL}. If jitter were not a consideration, $\sigma_{y, n}$ would fully capture the uncertainty in $y_n$, so we would expect
\begin{equation}
    \frac{1}{N-\gamma} \sum_{n=0}^{N-1} \Bigg(\frac{\Delta y_n - \tau_n}{\sigma_{y, n}}\Bigg)^2 \approx 1 .
\label{eq:red_chi2_without_jitter}
\end{equation}
Let us then define a normalized residual: 
\begin{equation}
     \delta_y \equiv [\Delta y_n - \tau_n]/\sigma_{y, n}.
     \label{eq:normresid}
\end{equation}
For a well-fitted model combined with reported uncertainties $\sigma_{y, n}$ that include all types of jitter, the distribution of $\delta_y$ should closely approximate a standard normal distribution $\mathcal{N}(0, 1)$. But since the $\sigma_{y, n}$ do not account for either instrumental or stellar jitter, we expect the $\delta_y$ distribution to be still roughly Gaussian, but broader than $\mathcal{N}(0, 1)$. 

We can therefore estimate the jitter for each instrument by calculating the sample variance of $\delta_y$. Fitting a Gaussian distribution $\mathcal{N}(0, \mathcal{W)}$ to the $\delta_y$ distribution for each instrument yields a weighting factor $\mathcal{W}$ that quantifies the degree to which $\sigma_{y}$ underestimates the uncertainty in $y_n$. Our estimator for the variance in $y_n$ is therefore:
\begin{equation}
    \widehat{\text{Var}}(y_n) = \sigma_{y, n}^2 + \sigma_{J, n}^2 \approx  (\mathcal{W}\sigma_{y, n})^2, 
\label{eq:true_var}
\end{equation}
which leads to a jitter estimate $\sigma_{J, n}$ at time $t_n$ of:
\begin{equation}
    \sigma_{J, n} \approx (\mathcal{W}^2 - 1)^{1/2} \sigma_{y, n} .
\label{eq:time_dep_jitter}
\end{equation}
We note that Equation \ref{eq:time_dep_jitter} does not fully capture the random nature of jitter, as the estimated $\sigma_{J, n}$ artificially depend on the reported $\sigma_{y,n}$ despite their different physical origins. However, Equation \ref{eq:time_dep_jitter} should yield a distribution of $\delta_y$ with approximately the same width as what would result from treating the stellar and instrumental jitter as free parameters.

\begin{figure}
    \centering
    \includegraphics[width=0.9\textwidth]{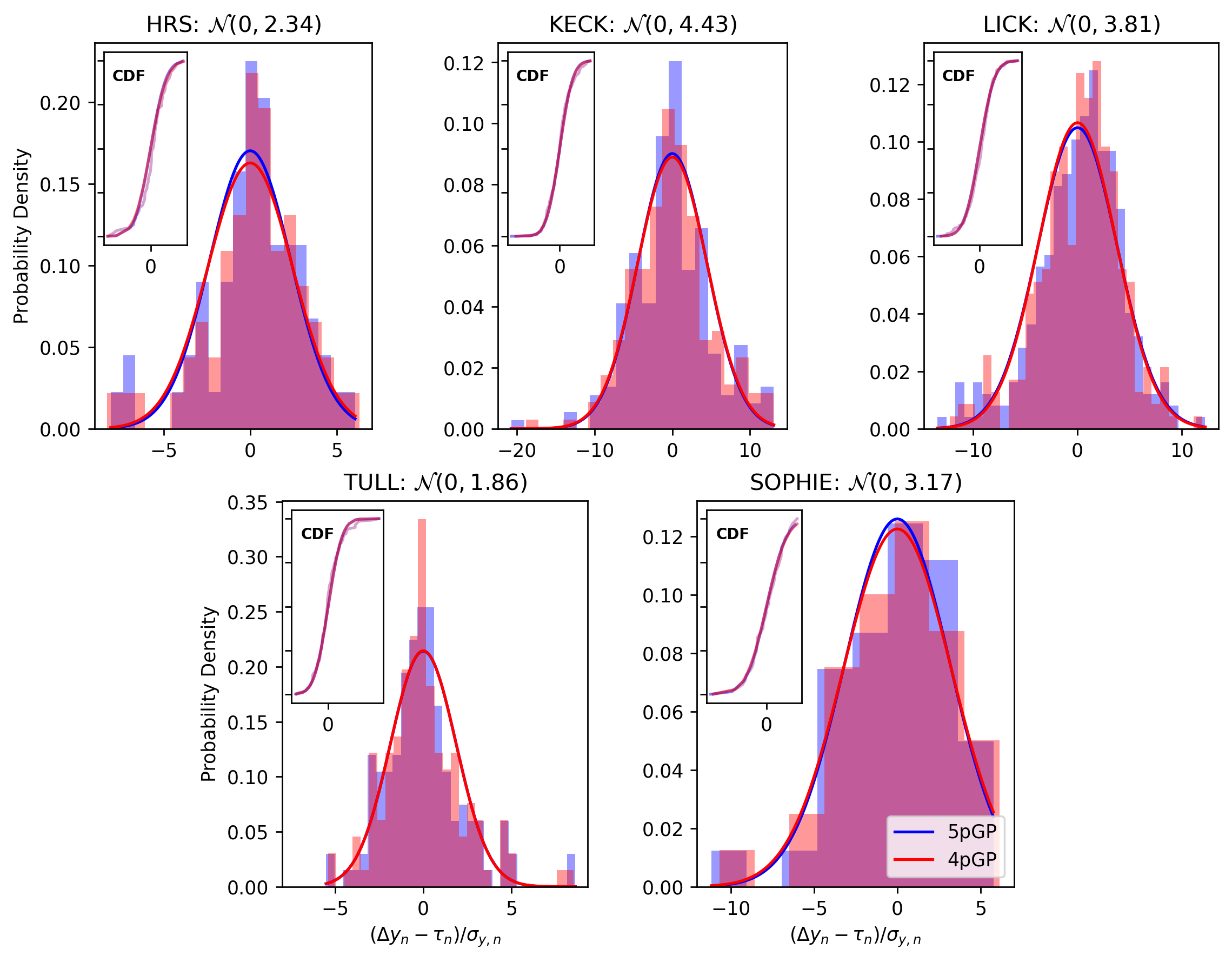}
    \caption{Distribution of residuals normalized by the reported observation uncertainty. The best-fit Gaussian distributions are superposed. We interpret the distributions being broader than a standard normal distribution $\mathcal{N}(0,1)$ as arising from jitter unaccounted for in the reported error bars. We estimate this combination of stellar and instrumental jitter via Equation \ref{eq:time_dep_jitter}. \textbf{Insets} - Cumulative distribution function fits for each normalized residual.}
    \label{fig:norm_res_distributions}
\end{figure}

For each instrument, we estimate $\mathcal{W}$ by fitting a Gaussian cumulative distribution function (CDF) to the empirical $\delta_y$ CDF. We exclude the HARPN and HARPS data, which have only 21 and 12 observations respectively, from CDF fitting as the small samples would lead to highly uncertain jitter estimates. We instead use a different approximation method for HARPN and HARPS jitter, explained below. Figure \ref{fig:norm_res_distributions} depicts the $\delta_y$ distributions and their empirical CDFs for each other instrument. 


To estimate jitter in the limited HARPN and HARPS data, we utilize an analytical formulation. Building from Equation \ref{eq:red_chi2_without_jitter}, the simplest approximation to instrument-specific jitter $\sigma_J$ (not dependent on $n$) assumes
\begin{equation}
    \frac{1}{N-\gamma} \sum_{n=0}^{N-1} (\Delta y_n - \tau_n)^2 \approx \bar{\sigma_y^2} + \sigma_J^2,
\label{eq:rough_jitter_approximation}
\end{equation}
with
\begin{equation}
    \bar{\sigma_y^2} = \frac{1}{N} \sum_{n=0}^{N-1} \sigma_{y, n}^2 .
\label{eq:mean_square_uncertainty}
\end{equation}
The estimated jitter comes from solving Equation \ref{eq:rough_jitter_approximation} for $\sigma_J$. 

Our jitter estimates for each instrument can be seen in Table \ref{table:instrument_fits}, which includes our best-fit RV offsets via Nelder-Mead optimization. We generally find good agreement between the jitter estimates that accompany the four-planet and five-planet models (4pGP and 5pGP, respectively). Our values are within $1-2$~ms$^{-1}$ of those published by \citet{Bourrier2018}, but not always within their uncertainties. Given that our treatment of jitter is approximate, we do not include uncertainties on our jitter estimates.

\bibliographystyle{aasjournal}
\typeout{}
\bibliography{biblio}

\begin{thebibliography}{}
\expandafter\ifx\csname natexlab\endcsname\relax\def\natexlab#1{#1}\fi
\providecommand{\url}[1]{\href{#1}{#1}}
\providecommand{\dodoi}[1]{doi:~\href{http://doi.org/#1}{\nolinkurl{#1}}}
\providecommand{\doeprint}[1]{\href{http://ascl.net/#1}{\nolinkurl{http://ascl.net/#1}}}
\providecommand{\doarXiv}[1]{\href{https://arxiv.org/abs/#1}{\nolinkurl{https://arxiv.org/abs/#1}}}

\bibitem[{{Aigrain} {et~al.}(2012){Aigrain}, {Pont}, \& {Zucker}}]{aigrain12}
{Aigrain}, S., {Pont}, F., \& {Zucker}, S. 2012, \mnras, 419, 3147, \dodoi{10.1111/j.1365-2966.2011.19960.x}

\bibitem[{{Alonso-Floriano} {et~al.}(2015){Alonso-Floriano}, {Morales}, {Caballero}, {Montes}, {Klutsch}, {Mundt}, {Cort{\'e}s-Contreras}, {Ribas}, {Reiners}, {Amado}, {Quirrenbach}, \& {Jeffers}}]{Alonso-Floriano2015}
{Alonso-Floriano}, F.~J., {Morales}, J.~C., {Caballero}, J.~A., {et~al.} 2015, \aap, 577, A128, \dodoi{10.1051/0004-6361/201525803}

\bibitem[{{Ambikasaran} {et~al.}(2015){Ambikasaran}, {Foreman-Mackey}, {Greengard}, {Hogg}, \& {O'Neil}}]{george_tech_paper}
{Ambikasaran}, S., {Foreman-Mackey}, D., {Greengard}, L., {Hogg}, D.~W., \& {O'Neil}, M. 2015, IEEE Transactions on Pattern Analysis and Machine Intelligence, 38, 252, \dodoi{10.1109/TPAMI.2015.2448083}

\bibitem[{Anderson \& Darling(1952)}]{Anderson_Darling1952}
Anderson, T.~W., \& Darling, D.~A. 1952, The Annals of Mathematical Statistics, 23, 193.
\newblock \url{http://www.jstor.org/stable/2236446}

\bibitem[{Anderson \& Darling(1954)}]{Anderson_Darling1954}
---. 1954, Journal of the American Statistical Association, 49, 765.
\newblock \url{http://www.jstor.org/stable/2281537}

\bibitem[{{Angus} {et~al.}(2018){Angus}, {Morton}, {Aigrain}, {Foreman-Mackey}, \& {Rajpaul}}]{Angus18}
{Angus}, R., {Morton}, T., {Aigrain}, S., {Foreman-Mackey}, D., \& {Rajpaul}, V. 2018, \mnras, 474, 2094, \dodoi{10.1093/mnras/stx2109}

\bibitem[{{Astropy Collaboration} {et~al.}(2013){Astropy Collaboration}, {Robitaille}, {Tollerud}, {Greenfield}, {Droettboom}, {Bray}, {Aldcroft}, {Davis}, {Ginsburg}, {Price-Whelan}, {Kerzendorf}, {Conley}, {Crighton}, {Barbary}, {Muna}, {Ferguson}, {Grollier}, {Parikh}, {Nair}, {Unther}, {Deil}, {Woillez}, {Conseil}, {Kramer}, {Turner}, {Singer}, {Fox}, {Weaver}, {Zabalza}, {Edwards}, {Azalee Bostroem}, {Burke}, {Casey}, {Crawford}, {Dencheva}, {Ely}, {Jenness}, {Labrie}, {Lim}, {Pierfederici}, {Pontzen}, {Ptak}, {Refsdal}, {Servillat}, \& {Streicher}}]{astropy:2013}
{Astropy Collaboration}, {Robitaille}, T.~P., {Tollerud}, E.~J., {et~al.} 2013, \aap, 558, A33, \dodoi{10.1051/0004-6361/201322068}

\bibitem[{{Astropy Collaboration} {et~al.}(2018){Astropy Collaboration}, {Price-Whelan}, {Sip{\H{o}}cz}, {G{\"u}nther}, {Lim}, {Crawford}, {Conseil}, {Shupe}, {Craig}, {Dencheva}, {Ginsburg}, {Vand erPlas}, {Bradley}, {P{\'e}rez-Su{\'a}rez}, {de Val-Borro}, {Aldcroft}, {Cruz}, {Robitaille}, {Tollerud}, {Ardelean}, {Babej}, {Bach}, {Bachetti}, {Bakanov}, {Bamford}, {Barentsen}, {Barmby}, {Baumbach}, {Berry}, {Biscani}, {Boquien}, {Bostroem}, {Bouma}, {Brammer}, {Bray}, {Breytenbach}, {Buddelmeijer}, {Burke}, {Calderone}, {Cano Rodr{\'\i}guez}, {Cara}, {Cardoso}, {Cheedella}, {Copin}, {Corrales}, {Crichton}, {D'Avella}, {Deil}, {Depagne}, {Dietrich}, {Donath}, {Droettboom}, {Earl}, {Erben}, {Fabbro}, {Ferreira}, {Finethy}, {Fox}, {Garrison}, {Gibbons}, {Goldstein}, {Gommers}, {Greco}, {Greenfield}, {Groener}, {Grollier}, {Hagen}, {Hirst}, {Homeier}, {Horton}, {Hosseinzadeh}, {Hu}, {Hunkeler}, {Ivezi{\'c}}, {Jain}, {Jenness}, {Kanarek}, {Kendrew}, {Kern}, {Kerzendorf}, {Khvalko}, {King}, {Kirkby}, {Kulkarni},
  {Kumar}, {Lee}, {Lenz}, {Littlefair}, {Ma}, {Macleod}, {Mastropietro}, {McCully}, {Montagnac}, {Morris}, {Mueller}, {Mumford}, {Muna}, {Murphy}, {Nelson}, {Nguyen}, {Ninan}, {N{\"o}the}, {Ogaz}, {Oh}, {Parejko}, {Parley}, {Pascual}, {Patil}, {Patil}, {Plunkett}, {Prochaska}, {Rastogi}, {Reddy Janga}, {Sabater}, {Sakurikar}, {Seifert}, {Sherbert}, {Sherwood-Taylor}, {Shih}, {Sick}, {Silbiger}, {Singanamalla}, {Singer}, {Sladen}, {Sooley}, {Sornarajah}, {Streicher}, {Teuben}, {Thomas}, {Tremblay}, {Turner}, {Terr{\'o}n}, {van Kerkwijk}, {de la Vega}, {Watkins}, {Weaver}, {Whitmore}, {Woillez}, {Zabalza}, \& {Astropy Contributors}}]{astropy:2018}
{Astropy Collaboration}, {Price-Whelan}, A.~M., {Sip{\H{o}}cz}, B.~M., {et~al.} 2018, \aj, 156, 123, \dodoi{10.3847/1538-3881/aabc4f}

\bibitem[{{Astropy Collaboration} {et~al.}(2022){Astropy Collaboration}, {Price-Whelan}, {Lim}, {Earl}, {Starkman}, {Bradley}, {Shupe}, {Patil}, {Corrales}, {Brasseur}, {N{"o}the}, {Donath}, {Tollerud}, {Morris}, {Ginsburg}, {Vaher}, {Weaver}, {Tocknell}, {Jamieson}, {van Kerkwijk}, {Robitaille}, {Merry}, {Bachetti}, {G{"u}nther}, {Aldcroft}, {Alvarado-Montes}, {Archibald}, {B{'o}di}, {Bapat}, {Barentsen}, {Baz{'a}n}, {Biswas}, {Boquien}, {Burke}, {Cara}, {Cara}, {Conroy}, {Conseil}, {Craig}, {Cross}, {Cruz}, {D'Eugenio}, {Dencheva}, {Devillepoix}, {Dietrich}, {Eigenbrot}, {Erben}, {Ferreira}, {Foreman-Mackey}, {Fox}, {Freij}, {Garg}, {Geda}, {Glattly}, {Gondhalekar}, {Gordon}, {Grant}, {Greenfield}, {Groener}, {Guest}, {Gurovich}, {Handberg}, {Hart}, {Hatfield-Dodds}, {Homeier}, {Hosseinzadeh}, {Jenness}, {Jones}, {Joseph}, {Kalmbach}, {Karamehmetoglu}, {Ka{l}uszy{'n}ski}, {Kelley}, {Kern}, {Kerzendorf}, {Koch}, {Kulumani}, {Lee}, {Ly}, {Ma}, {MacBride}, {Maljaars}, {Muna}, {Murphy}, {Norman}, {O'Steen},
  {Oman}, {Pacifici}, {Pascual}, {Pascual-Granado}, {Patil}, {Perren}, {Pickering}, {Rastogi}, {Roulston}, {Ryan}, {Rykoff}, {Sabater}, {Sakurikar}, {Salgado}, {Sanghi}, {Saunders}, {Savchenko}, {Schwardt}, {Seifert-Eckert}, {Shih}, {Jain}, {Shukla}, {Sick}, {Simpson}, {Singanamalla}, {Singer}, {Singhal}, {Sinha}, {Sip{H{o}}cz}, {Spitler}, {Stansby}, {Streicher}, {{{S}}umak}, {Swinbank}, {Taranu}, {Tewary}, {Tremblay}, {Val-Borro}, {Van Kooten}, {Vasovi{'c}}, {Verma}, {de Miranda Cardoso}, {Williams}, {Wilson}, {Winkel}, {Wood-Vasey}, {Xue}, {Yoachim}, {Zhang}, {Zonca}, \& {Astropy Project Contributors}}]{astropy:2022}
{Astropy Collaboration}, {Price-Whelan}, A.~M., {Lim}, P.~L., {et~al.} 2022, \apj, 935, 167, \dodoi{10.3847/1538-4357/ac7c74}

\bibitem[{{Baliunas} {et~al.}(1995){Baliunas}, {Donahue}, {Soon}, {Horne}, {Frazer}, {Woodard-Eklund}, {Bradford}, {Rao}, {Wilson}, {Zhang}, {Bennett}, {Briggs}, {Carroll}, {Duncan}, {Figueroa}, {Lanning}, {Misch}, {Mueller}, {Noyes}, {Poppe}, {Porter}, {Robinson}, {Russell}, {Shelton}, {Soyumer}, {Vaughan}, \& {Whitney}}]{Baliunas1995}
{Baliunas}, S.~L., {Donahue}, R.~A., {Soon}, W.~H., {et~al.} 1995, \apj, 438, 269, \dodoi{10.1086/175072}

\bibitem[{{Beard} {et~al.}(2024){Beard}, {Robertson}, {Dai}, {Holcomb}, {Lubin}, {Akana Murphy}, {Batalha}, {Blunt}, {Crossfield}, {Dressing}, {Fulton}, {Howard}, {Huber}, {Isaacson}, {Kane}, {Nowak}, {Petigura}, {Roy}, {Rubenzahl}, {Weiss}, {Barrena}, {Behmard}, {Brinkman}, {Carleo}, {Chontos}, {Dalba}, {Fetherolf}, {Giacalone}, {Hill}, {Kawauchi}, {Korth}, {Luque}, {MacDougall}, {Mayo}, {Mo{\v{c}}nik}, {Morello}, {Murgas}, {Orell-Miquel}, {Palle}, {Polanski}, {Rice}, {Scarsdale}, {Tyler}, \& {Van Zandt}}]{Beard2024}
{Beard}, C., {Robertson}, P., {Dai}, F., {et~al.} 2024, \aj, 167, 70, \dodoi{10.3847/1538-3881/ad1330}

\bibitem[{{Bevington} \& {Robinson}(2003)}]{bevington03}
{Bevington}, P.~R., \& {Robinson}, D.~K. 2003, {Data reduction and error analysis for the physical sciences}

\bibitem[{{Blunt} {et~al.}(2023){Blunt}, {Carvalho}, {David}, {Beichman}, {Zink}, {Gaidos}, {Behmard}, {Bouma}, {Cody}, {Dai}, {Foreman-Mackey}, {Grunblatt}, {Howard}, {Kosiarek}, {Knutson}, {Rubenzahl}, {Beard}, {Chontos}, {Giacalone}, {Hirano}, {Johnson}, {Lubin}, {Akana Murphy}, {Petigura}, {Van Zandt}, \& {Weiss}}]{Blunt2023}
{Blunt}, S., {Carvalho}, A., {David}, T.~J., {et~al.} 2023, \aj, 166, 62, \dodoi{10.3847/1538-3881/acde78}

\bibitem[{{Boisse} {et~al.}(2011){Boisse}, {Bouchy}, {H{\'e}brard}, {Bonfils}, {Santos}, \& {Vauclair}}]{Boisse2011}
{Boisse}, I., {Bouchy}, F., {H{\'e}brard}, G., {et~al.} 2011, \aap, 528, A4, \dodoi{10.1051/0004-6361/201014354}

\bibitem[{{Bortle} {et~al.}(2021){Bortle}, {Fausey}, {Ji}, {Dodson-Robinson}, {Ramirez Delgado}, \& {Gizis}}]{bortle21}
{Bortle}, A., {Fausey}, H., {Ji}, J., {et~al.} 2021, \aj, 161, 230, \dodoi{10.3847/1538-3881/abec89}

\bibitem[{{Bourrier} \& {H{\'e}brard}(2014)}]{Bourrier2014}
{Bourrier}, V., \& {H{\'e}brard}, G. 2014, \aap, 569, A65, \dodoi{10.1051/0004-6361/201424266}

\bibitem[{{Bourrier} {et~al.}(2018){Bourrier}, {Ehrenreich}, {Lecavelier Des Etangs}, {Louden}, {Wheatley}, {Wyttenbach}, {Vidal-Madjar}, {Lavie}, {Pepe}, \& {Udry}}]{Bourrier2018}
{Bourrier}, V., {Ehrenreich}, D., {Lecavelier Des Etangs}, A., {et~al.} 2018, VizieR Online Data Catalog, J/A+A/615/A117

\bibitem[{{Bramall} {et~al.}(2012){Bramall}, {Schmoll}, {Tyas}, {Clark}, {Younger}, {Sharples}, {Dipper}, {Ryan}, {Buckley}, \& {Brink}}]{Bramall2012}
{Bramall}, D.~G., {Schmoll}, J., {Tyas}, L.~M.~G., {et~al.} 2012, in Society of Photo-Optical Instrumentation Engineers (SPIE) Conference Series, Vol. 8446, Ground-based and Airborne Instrumentation for Astronomy IV, ed. I.~S. {McLean}, S.~K. {Ramsay}, \& H.~{Takami}, 84460A, \dodoi{10.1117/12.925935}

\bibitem[{{Bugnet}(2022)}]{Bugnet2022}
{Bugnet}, L. 2022, \aap, 667, A68, \dodoi{10.1051/0004-6361/202243167}

\bibitem[{{Butler} {et~al.}(1997){Butler}, {Marcy}, {Williams}, {Hauser}, \& {Shirts}}]{Butler1997}
{Butler}, R.~P., {Marcy}, G.~W., {Williams}, E., {Hauser}, H., \& {Shirts}, P. 1997, \apjl, 474, L115, \dodoi{10.1086/310444}

\bibitem[{{Butler} {et~al.}(2017){Butler}, {Vogt}, {Laughlin}, {Burt}, {Rivera}, {Tuomi}, {Teske}, {Arriagada}, {Diaz}, {Holden}, \& {Keiser}}]{Butler2017}
{Butler}, R.~P., {Vogt}, S.~S., {Laughlin}, G., {et~al.} 2017, \aj, 153, 208, \dodoi{10.3847/1538-3881/aa66ca}

\bibitem[{{Cabot} {et~al.}(2021){Cabot}, {Roettenbacher}, {Henry}, {Zhao}, {Harmon}, {Fischer}, {Brewer}, {Llama}, {Petersburg}, \& {Szymkowiak}}]{cabot21}
{Cabot}, S. H.~C., {Roettenbacher}, R.~M., {Henry}, G.~W., {et~al.} 2021, \aj, 161, 26, \dodoi{10.3847/1538-3881/abc41e}

\bibitem[{{Campante} {et~al.}(2011){Campante}, {Handberg}, {Mathur}, {Appourchaux}, {Bedding}, {Chaplin}, {Garc{\'\i}a}, {Mosser}, {Benomar}, {Bonanno}, {Corsaro}, {Fletcher}, {Gaulme}, {Hekker}, {Karoff}, {R{\'e}gulo}, {Salabert}, {Verner}, {White}, {Houdek}, {Brand{\~a}o}, {Creevey}, {Do{\v{g}}an}, {Bazot}, {Christensen-Dalsgaard}, {Cunha}, {Elsworth}, {Huber}, {Kjeldsen}, {Lundkvist}, {Molenda-{\.Z}akowicz}, {Monteiro}, {Stello}, {Clarke}, {Girouard}, \& {Hall}}]{Campante2011}
{Campante}, T.~L., {Handberg}, R., {Mathur}, S., {et~al.} 2011, \aap, 534, A6, \dodoi{10.1051/0004-6361/201116620}

\bibitem[{{Carolo} {et~al.}(2014){Carolo}, {Desidera}, {Gratton}, {Martinez Fiorenzano}, {Marzari}, {Endl}, {Mesa}, {Barbieri}, {Cecconi}, {Claudi}, {Cosentino}, \& {Scuderi}}]{Carolo2014}
{Carolo}, E., {Desidera}, S., {Gratton}, R., {et~al.} 2014, \aap, 567, A48, \dodoi{10.1051/0004-6361/201323102}

\bibitem[{{Cegla}(2019)}]{Cegla2019}
{Cegla}, H.~M. 2019, Geosciences, 9, 114, \dodoi{10.3390/geosciences9030114}

\bibitem[{{Chowdhury} {et~al.}(2019){Chowdhury}, {Kilcik}, {Yurchyshyn}, {Obridko}, \& {Rozelot}}]{Chowdhury2019}
{Chowdhury}, P., {Kilcik}, A., {Yurchyshyn}, V., {Obridko}, V.~N., \& {Rozelot}, J.~P. 2019, \solphys, 294, 142, \dodoi{10.1007/s11207-019-1530-7}

\bibitem[{{Cosentino} {et~al.}(2012){Cosentino}, {Lovis}, {Pepe}, {Collier Cameron}, {Latham}, {Molinari}, {Udry}, {Bezawada}, {Black}, {Born}, {Buchschacher}, {Charbonneau}, {Figueira}, {Fleury}, {Galli}, {Gallie}, {Gao}, {Ghedina}, {Gonzalez}, {Gonzalez}, {Guerra}, {Henry}, {Horne}, {Hughes}, {Kelly}, {Lodi}, {Lunney}, {Maire}, {Mayor}, {Micela}, {Ordway}, {Peacock}, {Phillips}, {Piotto}, {Pollacco}, {Queloz}, {Rice}, {Riverol}, {Riverol}, {San Juan}, {Sasselov}, {Segransan}, {Sozzetti}, {Sosnowska}, {Stobie}, {Szentgyorgyi}, {Vick}, \& {Weber}}]{Cosentino2012}
{Cosentino}, R., {Lovis}, C., {Pepe}, F., {et~al.} 2012, in Society of Photo-Optical Instrumentation Engineers (SPIE) Conference Series, Vol. 8446, Ground-based and Airborne Instrumentation for Astronomy IV, ed. I.~S. {McLean}, S.~K. {Ramsay}, \& H.~{Takami}, 84461V, \dodoi{10.1117/12.925738}

\bibitem[{{Dawson} \& {Fabrycky}(2010)}]{Dawson2010}
{Dawson}, R.~I., \& {Fabrycky}, D.~C. 2010, \apj, 722, 937, \dodoi{10.1088/0004-637X/722/1/937}

\bibitem[{{Delisle} {et~al.}(2018){Delisle}, {S{\'e}gransan}, {Dumusque}, {Diaz}, {Bouchy}, {Lovis}, {Pepe}, {Udry}, {Alonso}, {Benz}, {Coffinet}, {Collier Cameron}, {Deleuil}, {Figueira}, {Gillon}, {Lo Curto}, {Mayor}, {Mordasini}, {Motalebi}, {Moutou}, {Pollacco}, {Pompei}, {Queloz}, {Santos}, \& {Wyttenbach}}]{delisle18}
{Delisle}, J.~B., {S{\'e}gransan}, D., {Dumusque}, X., {et~al.} 2018, \aap, 614, A133, \dodoi{10.1051/0004-6361/201732529}

\bibitem[{{Demory} {et~al.}(2011){Demory}, {Gillon}, {Deming}, \& {Seager}}]{Demory2011}
{Demory}, B.-O., {Gillon}, M., {Deming}, D., \& {Seager}, S. 2011, in AAS/Division for Extreme Solar Systems Abstracts, Vol.~2, AAS/Division for Extreme Solar Systems Abstracts, 17.04

\bibitem[{{Demory} {et~al.}(2016{\natexlab{a}}){Demory}, {Gillon}, {Madhusudhan}, \& {Queloz}}]{Demory2016B}
{Demory}, B.-O., {Gillon}, M., {Madhusudhan}, N., \& {Queloz}, D. 2016{\natexlab{a}}, \mnras, 455, 2018, \dodoi{10.1093/mnras/stv2239}

\bibitem[{{Demory} {et~al.}(2016{\natexlab{b}}){Demory}, {Gillon}, {de Wit}, {Madhusudhan}, {Bolmont}, {Heng}, {Kataria}, {Lewis}, {Hu}, {Krick}, {Stamenkovi{\'c}}, {Benneke}, {Kane}, \& {Queloz}}]{Demory2016A}
{Demory}, B.-O., {Gillon}, M., {de Wit}, J., {et~al.} 2016{\natexlab{b}}, \nat, 532, 207, \dodoi{10.1038/nature17169}

\bibitem[{{Desidera} {et~al.}(2004){Desidera}, {Gratton}, {Endl}, {Claudi}, {Cosentino}, {Barbieri}, {Bonanno}, {Lucatello}, {Martinez Fiorenzano}, {Marzari}, \& {Scuderi}}]{Desidera2004}
{Desidera}, S., {Gratton}, R.~G., {Endl}, M., {et~al.} 2004, \aap, 420, L27, \dodoi{10.1051/0004-6361:20040155}

\bibitem[{{Desort} {et~al.}(2007){Desort}, {Lagrange}, {Galland}, {Udry}, \& {Mayor}}]{Desort2007}
{Desort}, M., {Lagrange}, A.~M., {Galland}, F., {Udry}, S., \& {Mayor}, M. 2007, \aap, 473, 983, \dodoi{10.1051/0004-6361:20078144}

\bibitem[{{Dodson-Robinson} \& {Haley}(2024)}]{SDR_Haley2024}
{Dodson-Robinson}, S., \& {Haley}, C. 2024, \aj, 167, 22, \dodoi{10.3847/1538-3881/ad0c58}

\bibitem[{{Dodson-Robinson} {et~al.}(2022){Dodson-Robinson}, {Delgado}, {Harrell}, \& {Haley}}]{SDR2022}
{Dodson-Robinson}, S.~E., {Delgado}, V.~R., {Harrell}, J., \& {Haley}, C.~L. 2022, \aj, 163, 169, \dodoi{10.3847/1538-3881/ac52ed}

\bibitem[{{Dumusque}(2012)}]{Dumusque2012}
{Dumusque}, X. 2012, PhD thesis, University of Geneva, Astronomical Observatory

\bibitem[{{Dumusque} {et~al.}(2011){Dumusque}, {Udry}, {Lovis}, {Santos}, \& {Monteiro}}]{Dumusque2011}
{Dumusque}, X., {Udry}, S., {Lovis}, C., {Santos}, N.~C., \& {Monteiro}, M.~J.~P.~F.~G. 2011, \aap, 525, A140, \dodoi{10.1051/0004-6361/201014097}

\bibitem[{{Dumusque} {et~al.}(2017){Dumusque}, {Borsa}, {Damasso}, {D{\'\i}az}, {Gregory}, {Hara}, {Hatzes}, {Rajpaul}, {Tuomi}, {Aigrain}, {Anglada-Escud{\'e}}, {Bonomo}, {Bou{\'e}}, {Dauvergne}, {Frustagli}, {Giacobbe}, {Haywood}, {Jones}, {Laskar}, {Pinamonti}, {Poretti}, {Rainer}, {S{\'e}gransan}, {Sozzetti}, \& {Udry}}]{Dumusque2017}
{Dumusque}, X., {Borsa}, F., {Damasso}, M., {et~al.} 2017, \aap, 598, A133, \dodoi{10.1051/0004-6361/201628671}

\bibitem[{{Ehrenreich} {et~al.}(2012){Ehrenreich}, {Bourrier}, {Bonfils}, {Lecavelier des Etangs}, {H{\'e}brard}, {Sing}, {Wheatley}, {Vidal-Madjar}, {Delfosse}, {Udry}, {Forveille}, \& {Moutou}}]{Ehrenreich2012}
{Ehrenreich}, D., {Bourrier}, V., {Bonfils}, X., {et~al.} 2012, \aap, 547, A18, \dodoi{10.1051/0004-6361/201219981}

\bibitem[{{El-Borie} {et~al.}(2020){El-Borie}, {El-Taher}, {Thabet}, {Ibrahim}, {Aly}, \& {Bishara}}]{El-Borie2020}
{El-Borie}, M.~A., {El-Taher}, A.~M., {Thabet}, A.~A., {et~al.} 2020, \apj, 898, 73, \dodoi{10.3847/1538-4357/ab9d21}

\bibitem[{{Endl} {et~al.}(2012){Endl}, {Robertson}, {Cochran}, {MacQueen}, {Brugamyer}, {Caldwell}, {Wittenmyer}, {Barnes}, \& {Gullikson}}]{Endl2012}
{Endl}, M., {Robertson}, P., {Cochran}, W.~D., {et~al.} 2012, \apj, 759, 19, \dodoi{10.1088/0004-637X/759/1/19}

\bibitem[{{Esteves} {et~al.}(2017){Esteves}, {de Mooij}, {Jayawardhana}, {Watson}, \& {de Kok}}]{Esteves2017}
{Esteves}, L.~J., {de Mooij}, E. J.~W., {Jayawardhana}, R., {Watson}, C., \& {de Kok}, R. 2017, \aj, 153, 268, \dodoi{10.3847/1538-3881/aa7133}

\bibitem[{{Farr} {et~al.}(2018){Farr}, {Pope}, {Davies}, {North}, {White}, {Barrett}, {Miglio}, {Lund}, {Antoci}, {Fredslund Andersen}, {Grundahl}, \& {Huber}}]{farr18}
{Farr}, W.~M., {Pope}, B. J.~S., {Davies}, G.~R., {et~al.} 2018, \apjl, 865, L20, \dodoi{10.3847/2041-8213/aadfde}

\bibitem[{{Fischer} {et~al.}(2008){Fischer}, {Marcy}, {Butler}, {Vogt}, {Laughlin}, {Henry}, {Abouav}, {Peek}, {Wright}, {Johnson}, {McCarthy}, \& {Isaacson}}]{Fischer2008}
{Fischer}, D.~A., {Marcy}, G.~W., {Butler}, R.~P., {et~al.} 2008, \apj, 675, 790, \dodoi{10.1086/525512}

\bibitem[{{Fischer} {et~al.}(2016){Fischer}, {Anglada-Escude}, {Arriagada}, {Baluev}, {Bean}, {Bouchy}, {Buchhave}, {Carroll}, {Chakraborty}, {Crepp}, {Dawson}, {Diddams}, {Dumusque}, {Eastman}, {Endl}, {Figueira}, {Ford}, {Foreman-Mackey}, {Fournier}, {F{\H{u}}r{\'e}sz}, {Gaudi}, {Gregory}, {Grundahl}, {Hatzes}, {H{\'e}brard}, {Herrero}, {Hogg}, {Howard}, {Johnson}, {Jorden}, {Jurgenson}, {Latham}, {Laughlin}, {Loredo}, {Lovis}, {Mahadevan}, {McCracken}, {Pepe}, {Perez}, {Phillips}, {Plavchan}, {Prato}, {Quirrenbach}, {Reiners}, {Robertson}, {Santos}, {Sawyer}, {Segransan}, {Sozzetti}, {Steinmetz}, {Szentgyorgyi}, {Udry}, {Valenti}, {Wang}, {Wittenmyer}, \& {Wright}}]{Fischer2016}
{Fischer}, D.~A., {Anglada-Escude}, G., {Arriagada}, P., {et~al.} 2016, \pasp, 128, 066001, \dodoi{10.1088/1538-3873/128/964/066001}

\bibitem[{Foreman-Mackey {et~al.}(2017)Foreman-Mackey, Agol, Ambikasaran, \& Angus}]{Foreman-Mackey_2017}
Foreman-Mackey, D., Agol, E., Ambikasaran, S., \& Angus, R. 2017, The Astronomical Journal, 154, 220, \dodoi{10.3847/1538-3881/aa9332}

\bibitem[{{Fuhrmeister} {et~al.}(2023{\natexlab{a}}){Fuhrmeister}, {Coffaro}, {Stelzer}, {Mittag}, {Czesla}, \& {Schneider}}]{Fuhrmeister2023}
{Fuhrmeister}, B., {Coffaro}, M., {Stelzer}, B., {et~al.} 2023{\natexlab{a}}, \aap, 672, A149, \dodoi{10.1051/0004-6361/202245201}

\bibitem[{{Fuhrmeister} {et~al.}(2023{\natexlab{b}}){Fuhrmeister}, {Coffaro}, {Stelzer}, {Mittag}, {Czesla}, \& {Schneider}}]{fuhrmeister23}
---. 2023{\natexlab{b}}, \aap, 672, A149, \dodoi{10.1051/0004-6361/202245201}

\bibitem[{Gao \& Han(2012)}]{Gao2012}
Gao, F., \& Han, L. 2012, Computational Optimization and Applications, 51, 259 , \dodoi{10.1007/s10589-010-9329-3}

\bibitem[{{Godin}(1972)}]{Godin1972}
{Godin}, G. 1972, {The analysis of tides.} (Toronto: Univ. of Toronto Press)

\bibitem[{{Gomes da Silva} {et~al.}(2012){Gomes da Silva}, {Santos}, {Bonfils}, {Delfosse}, {Forveille}, {Udry}, {Dumusque}, \& {Lovis}}]{GomesdaSilva2012}
{Gomes da Silva}, J., {Santos}, N.~C., {Bonfils}, X., {et~al.} 2012, \aap, 541, A9, \dodoi{10.1051/0004-6361/201118598}

\bibitem[{Gray(2009)}]{Gray2009}
Gray, D.~F. 2009, The Astrophysical Journal, 697, 1032, \dodoi{10.1088/0004-637X/697/2/1032}

\bibitem[{Green \& Silverman(1994)}]{curvature_penalty_textbook}
Green, P., \& Silverman, B. 1994, Nonparametric regression and generalized linear models: a roughness penalty approach (United Kingdom: Chapman and Hall)

\bibitem[{Harrell(2024)}]{harrell_2024_55Cnc_code}
Harrell, J. 2024, 55Cnc\_exoplanets, v1.0.0,  Zenodo, \dodoi{10.5281/zenodo.14571275}

\bibitem[{Hathaway(2015)}]{hathaway15}
Hathaway, D.~H. 2015, Living Reviews in Solar Physics, 12, 4, \dodoi{10.1007/lrsp-2015-4}

\bibitem[{{Hatzes}(2002)}]{Hatzes2002}
{Hatzes}, A.~P. 2002, Astronomische Nachrichten, 323, 392, \dodoi{10.1002/1521-3994(200208)323:3/4<392::AID-ASNA392>3.0.CO;2-M}

\bibitem[{Hatzes(2019)}]{Hatzes2019}
Hatzes, A.~P. 2019, The Doppler Method for the Detection of Exoplanets, 2514-3433 (IOP Publishing), \dodoi{10.1088/2514-3433/ab46a3}

\bibitem[{{Haywood} {et~al.}(2014){Haywood}, {Collier Cameron}, {Queloz}, {Barros}, {Deleuil}, {Fares}, {Gillon}, {Lanza}, {Lovis}, {Moutou}, {Pepe}, {Pollacco}, {Santerne}, {S{\'e}gransan}, \& {Unruh}}]{Haywood2014}
{Haywood}, R.~D., {Collier Cameron}, A., {Queloz}, D., {et~al.} 2014, \mnras, 443, 2517, \dodoi{10.1093/mnras/stu1320}

\bibitem[{{Hinshaw} {et~al.}(2003){Hinshaw}, {Barnes}, {Bennett}, {Greason}, {Halpern}, {Hill}, {Jarosik}, {Kogut}, {Limon}, {Meyer}, {Odegard}, {Page}, {Spergel}, {Tucker}, {Weiland}, {Wollack}, \& {Wright}}]{Hinshaw2003}
{Hinshaw}, G., {Barnes}, C., {Bennett}, C.~L., {et~al.} 2003, \apjs, 148, 63, \dodoi{10.1086/377222}

\bibitem[{Hodrick \& Prescott(1997)}]{HP1997}
Hodrick, R.~J., \& Prescott, E.~C. 1997, Journal of Money, Credit and Banking, 29, 1.
\newblock \url{http://www.jstor.org/stable/2953682}

\bibitem[{Kass \& Raftery(1995)}]{Delta_BIC_10}
Kass, R.~E., \& Raftery, A.~E. 1995, Journal of the American Statistical Association, 90, 773, \dodoi{10.1080/01621459.1995.10476572}

\bibitem[{{Kervella} {et~al.}(2019){Kervella}, {Arenou}, {Mignard}, \& {Th{\'e}venin}}]{kervella19}
{Kervella}, P., {Arenou}, F., {Mignard}, F., \& {Th{\'e}venin}, F. 2019, \aap, 623, A72, \dodoi{10.1051/0004-6361/201834371}

\bibitem[{{L{\'o}pez-Morales} {et~al.}(2014){L{\'o}pez-Morales}, {Triaud}, {Rodler}, {Dumusque}, {Buchhave}, {Harutyunyan}, {Hoyer}, {Alonso}, {Gillon}, {Kaib}, {Latham}, {Lovis}, {Pepe}, {Queloz}, {Raymond}, {S{\'e}gransan}, {Waldmann}, \& {Udry}}]{Lopez2014}
{L{\'o}pez-Morales}, M., {Triaud}, A. H.~M.~J., {Rodler}, F., {et~al.} 2014, \apjl, 792, L31, \dodoi{10.1088/2041-8205/792/2/L31}

\bibitem[{{Lovis} \& {Fischer}(2010)}]{Lovis_Fischer2010}
{Lovis}, C., \& {Fischer}, D. 2010, in Exoplanets, ed. S.~{Seager} (University of Arizona Press), 27--53

\bibitem[{{Lubin} {et~al.}(2021){Lubin}, {Robertson}, {Stefansson}, {Ninan}, {Mahadevan}, {Endl}, {Ford}, {Wright}, {Beard}, {Bender}, {Cochran}, {Diddams}, {Fredrick}, {Halverson}, {Kanodia}, {Metcalf}, {Ramsey}, {Roy}, {Schwab}, \& {Terrien}}]{lubin21}
{Lubin}, J., {Robertson}, P., {Stefansson}, G., {et~al.} 2021, \aj, 162, 61, \dodoi{10.3847/1538-3881/ac0057}

\bibitem[{{Marcy} {et~al.}(2002){Marcy}, {Butler}, {Fischer}, {Laughlin}, {Vogt}, {Henry}, \& {Pourbaix}}]{Marcy2002}
{Marcy}, G.~W., {Butler}, R.~P., {Fischer}, D.~A., {et~al.} 2002, \apj, 581, 1375, \dodoi{10.1086/344298}

\bibitem[{{McArthur} {et~al.}(2004){McArthur}, {Endl}, {Cochran}, {Benedict}, {Fischer}, {Marcy}, {Butler}, {Naef}, {Mayor}, {Queloz}, {Udry}, \& {Harrison}}]{McArthur2004}
{McArthur}, B.~E., {Endl}, M., {Cochran}, W.~D., {et~al.} 2004, \apjl, 614, L81, \dodoi{10.1086/425561}

\bibitem[{McClellan \& Schafer(1999)}]{McClellan1999}
McClellan, J., \& Schafer, R. 1999, IEEE Signal Processing Magazine, 16, 29, \dodoi{10.1109/MSP.1999.790977}

\bibitem[{{Meunier} {et~al.}(2010){Meunier}, {Desort}, \& {Lagrange}}]{Meunier2010}
{Meunier}, N., {Desort}, M., \& {Lagrange}, A.~M. 2010, \aap, 512, A39, \dodoi{10.1051/0004-6361/200913551}

\bibitem[{Meunier \& Lagrange(2013)}]{Meunier2013}
Meunier, N., \& Lagrange, A. 2013, Astronomische Nachrichten, 334, 141, \dodoi{https://doi.org/10.1002/asna.201211742}

\bibitem[{Nelder \& Mead(1965)}]{NM_1965}
Nelder, J.~A., \& Mead, R. 1965, The Computer Journal, 7, 308, \dodoi{10.1093/comjnl/7.4.308}

\bibitem[{Newton {et~al.}(2016)Newton, Irwin, Charbonneau, Berta-Thompson, \& Dittmann}]{Newton2016}
Newton, E.~R., Irwin, J., Charbonneau, D., Berta-Thompson, Z.~K., \& Dittmann, J.~A. 2016, The Astrophysical Journal Letters, 821, L19, \dodoi{10.3847/2041-8205/821/1/L19}

\bibitem[{Ochsenbein(1996)}]{10.26093/cds/vizier}
Ochsenbein, F. 1996, The VizieR database of astronomical catalogues,  CDS, Centre de Données astronomiques de Strasbourg, \dodoi{10.26093/CDS/VIZIER}

\bibitem[{{Ochsenbein} {et~al.}(2000){Ochsenbein}, {Bauer}, \& {Marcout}}]{vizier_origin_paper}
{Ochsenbein}, F., {Bauer}, P., \& {Marcout}, J. 2000, \aaps, 143, 23, \dodoi{10.1051/aas:2000169}

\bibitem[{{Pepe} {et~al.}(2003){Pepe}, {Rupprecht}, {Avila}, {Balestra}, {Bouchy}, {Cavadore}, {Eckert}, {Fleury}, {Gillotte}, {Gojak}, {Guzman}, {Kohler}, {Lizon}, {Mayor}, {Megevand}, {Queloz}, {Sosnowska}, {Udry}, \& {Weilenmann}}]{Pepe2003}
{Pepe}, F., {Rupprecht}, G., {Avila}, G., {et~al.} 2003, in Society of Photo-Optical Instrumentation Engineers (SPIE) Conference Series, Vol. 4841, Instrument Design and Performance for Optical/Infrared Ground-based Telescopes, ed. M.~{Iye} \& A.~F.~M. {Moorwood}, 1045--1056, \dodoi{10.1117/12.460777}

\bibitem[{{Perruchot} {et~al.}(2008){Perruchot}, {Kohler}, {Bouchy}, {Richaud}, {Richaud}, {Moreaux}, {Merzougui}, {Sottile}, {Hill}, {Knispel}, {Regal}, {Meunier}, {Ilovaisky}, {Le Coroller}, {Gillet}, {Schmitt}, {Pepe}, {Fleury}, {Sosnowska}, {Vors}, {M{\'e}gevand}, {Blanc}, {Carol}, {Point}, {Laloge}, \& {Brunel}}]{Perruchot2008}
{Perruchot}, S., {Kohler}, D., {Bouchy}, F., {et~al.} 2008, in Society of Photo-Optical Instrumentation Engineers (SPIE) Conference Series, Vol. 7014, Ground-based and Airborne Instrumentation for Astronomy II, ed. I.~S. {McLean} \& M.~M. {Casali}, 70140J, \dodoi{10.1117/12.787379}

\bibitem[{{Perryman} {et~al.}(2014){Perryman}, {Hartman}, {Bakos}, \& {Lindegren}}]{Perryman2014}
{Perryman}, M., {Hartman}, J., {Bakos}, G.~{\'A}., \& {Lindegren}, L. 2014, \apj, 797, 14, \dodoi{10.1088/0004-637X/797/1/14}

\bibitem[{{Rajpaul} {et~al.}(2015){Rajpaul}, {Aigrain}, {Osborne}, {Reece}, \& {Roberts}}]{Rajpaul2015}
{Rajpaul}, V., {Aigrain}, S., {Osborne}, M.~A., {Reece}, S., \& {Roberts}, S. 2015, \mnras, 452, 2269, \dodoi{10.1093/mnras/stv1428}

\bibitem[{{Rajpaul} {et~al.}(2021){Rajpaul}, {Buchhave}, {Lacedelli}, {Rice}, {Mortier}, {Malavolta}, {Aigrain}, {Borsato}, {Mayo}, {Charbonneau}, {Damasso}, {Dumusque}, {Ghedina}, {Latham}, {L{\'o}pez-Morales}, {Magazz{\`u}}, {Micela}, {Molinari}, {Pepe}, {Piotto}, {Poretti}, {Rowther}, {Sozzetti}, {Udry}, \& {Watson}}]{Rajpaul2021}
{Rajpaul}, V.~M., {Buchhave}, L.~A., {Lacedelli}, G., {et~al.} 2021, \mnras, 507, 1847, \dodoi{10.1093/mnras/stab2192}

\bibitem[{{Ramirez Delgado} {et~al.}(2025){Ramirez Delgado}, {Caicedo Vivas}, {Dodson-Robinson}, \& {Haley}}]{ramirezdelgado25}
{Ramirez Delgado}, V., {Caicedo Vivas}, J.~S., {Dodson-Robinson}, S., \& {Haley}, C. 2025, arXiv e-prints, arXiv:2506.20864, \dodoi{10.48550/arXiv.2506.20864}

\bibitem[{Rasmussen \& Williams(2005)}]{Rasmussen_GP_textbook}
Rasmussen, C.~E., \& Williams, C. K.~I. 2005, {Gaussian Processes for Machine Learning} (The MIT Press), \dodoi{10.7551/mitpress/3206.001.0001}

\bibitem[{Ravn \& Uhlig(2002)}]{Morten_adjusting_HP_filter}
Ravn, M.~O., \& Uhlig, H. 2002, The Review of Economics and Statistics, 84, 371, \dodoi{10.1162/003465302317411604}

\bibitem[{{Ridden-Harper} {et~al.}(2016){Ridden-Harper}, {Snellen}, {Keller}, {de Kok}, {Di Gloria}, {Hoeijmakers}, {Brogi}, {Fridlund}, {Vermeersen}, \& {van Westrenen}}]{Ridden-Harper2016}
{Ridden-Harper}, A.~R., {Snellen}, I.~A.~G., {Keller}, C.~U., {et~al.} 2016, \aap, 593, A129, \dodoi{10.1051/0004-6361/201628448}

\bibitem[{{Robertson} {et~al.}(2013{\natexlab{a}}){Robertson}, {Endl}, {Cochran}, \& {Dodson-Robinson}}]{Robertson2013}
{Robertson}, P., {Endl}, M., {Cochran}, W.~D., \& {Dodson-Robinson}, S.~E. 2013{\natexlab{a}}, \apj, 764, 3, \dodoi{10.1088/0004-637X/764/1/3}

\bibitem[{{Robertson} {et~al.}(2013{\natexlab{b}}){Robertson}, {Endl}, {Cochran}, \& {Dodson-Robinson}}]{robertson13}
---. 2013{\natexlab{b}}, \apj, 764, 3, \dodoi{10.1088/0004-637X/764/1/3}

\bibitem[{Robertson \& Mahadevan(2014)}]{Robertson2014_A}
Robertson, P., \& Mahadevan, S. 2014, The Astrophysical Journal Letters, 793, L24, \dodoi{10.1088/2041-8205/793/2/L24}

\bibitem[{{Robertson} {et~al.}(2014){Robertson}, {Mahadevan}, {Endl}, \& {Roy}}]{Robertson2014_B}
{Robertson}, P., {Mahadevan}, S., {Endl}, M., \& {Roy}, A. 2014, Science, 345, 440, \dodoi{10.1126/science.1253253}

\bibitem[{{Saar} \& {Donahue}(1997)}]{Saar1997}
{Saar}, S.~H., \& {Donahue}, R.~A. 1997, \apj, 485, 319, \dodoi{10.1086/304392}

\bibitem[{Scholz \& Stephens(1987)}]{Scholz1987}
Scholz, F.~W., \& Stephens, M.~A. 1987, Journal of the American Statistical Association, 82, 918.
\newblock \url{http://www.jstor.org/stable/2288805}

\bibitem[{{Schwarz}(1978)}]{Schwarz1978}
{Schwarz}, G. 1978, Annals of Statistics, 6, 461

\bibitem[{Shumway \& Stoffer(2001)}]{shumwaystoffer}
Shumway, R.~H., \& Stoffer, D.~S. 2001, Time Series Analysis and Its Applications, 4th edn. (New York: Springer)

\bibitem[{{Su{\'a}rez Mascare{\~n}o} {et~al.}(2016){Su{\'a}rez Mascare{\~n}o}, {Rebolo}, \& {Gonz{\'a}lez Hern{\'a}ndez}}]{suarezmascareno16}
{Su{\'a}rez Mascare{\~n}o}, A., {Rebolo}, R., \& {Gonz{\'a}lez Hern{\'a}ndez}, J.~I. 2016, \aap, 595, A12, \dodoi{10.1051/0004-6361/201628586}

\bibitem[{{Su{\'a}rez Mascare{\~n}o} {et~al.}(2017){Su{\'a}rez Mascare{\~n}o}, {Rebolo}, {Gonz{\'a}lez Hern{\'a}ndez}, \& {Esposito}}]{SuarezMascareno2017}
{Su{\'a}rez Mascare{\~n}o}, A., {Rebolo}, R., {Gonz{\'a}lez Hern{\'a}ndez}, J.~I., \& {Esposito}, M. 2017, \mnras, 468, 4772, \dodoi{10.1093/mnras/stx771}

\bibitem[{{Su{\'a}rez Mascare{\~n}o} {et~al.}(2023){Su{\'a}rez Mascare{\~n}o}, {Gonz{\'a}lez-{\'A}lvarez}, {Zapatero Osorio}, {Lillo-Box}, {Faria}, {Passegger}, {Gonz{\'a}lez Hern{\'a}ndez}, {Figueira}, {Sozzetti}, {Rebolo}, {Pepe}, {Santos}, {Cristiani}, {Lovis}, {Silva}, {Ribas}, {Amado}, {Caballero}, {Quirrenbach}, {Reiners}, {Zechmeister}, {Adibekyan}, {Alibert}, {B{\'e}jar}, {Benatti}, {D'Odorico}, {Damasso}, {Delisle}, {Di Marcantonio}, {Dreizler}, {Ehrenreich}, {Hatzes}, {Hara}, {Henning}, {Kaminski}, {L{\'o}pez-Gonz{\'a}lez}, {Martins}, {Micela}, {Montes}, {Pall{\'e}}, {Pedraz}, {Rodr{\'\i}guez}, {Rodr{\'\i}guez-L{\'o}pez}, {Tal-Or}, {Sousa}, \& {Udry}}]{suarezmascareno23}
{Su{\'a}rez Mascare{\~n}o}, A., {Gonz{\'a}lez-{\'A}lvarez}, E., {Zapatero Osorio}, M.~R., {et~al.} 2023, \aap, 670, A5, \dodoi{10.1051/0004-6361/202244991}

\bibitem[{{Thomson} \& {Haley}(2014)}]{Haley2014}
{Thomson}, D.~J., \& {Haley}, C.~L. 2014, Proceedings of the Royal Society of London Series A, 470, 20140101, \dodoi{10.1098/rspa.2014.0101}

\bibitem[{{Tull} {et~al.}(1995){Tull}, {MacQueen}, {Sneden}, \& {Lambert}}]{Tull1995}
{Tull}, R.~G., {MacQueen}, P.~J., {Sneden}, C., \& {Lambert}, D.~L. 1995, \pasp, 107, 251, \dodoi{10.1086/133548}

\bibitem[{Virtanen {et~al.}(2020)Virtanen, Gommers, Oliphant, Haberland, Reddy, Cournapeau, Burovski, Peterson, Weckesser, Bright, {van der Walt}, Brett, Wilson, Millman, Mayorov, Nelson, Jones, Kern, Larson, Carey, Polat, Feng, Moore, {VanderPlas}, Laxalde, Perktold, Cimrman, Henriksen, Quintero, Harris, Archibald, Ribeiro, Pedregosa, {van Mulbregt}, \& {SciPy 1.0 Contributors}}]{SciPy-NMeth2020}
Virtanen, P., Gommers, R., Oliphant, T.~E., {et~al.} 2020, Nature Methods, 17, 261, \dodoi{10.1038/s41592-019-0686-2}

\bibitem[{{Vogt}(1987)}]{Vogt1987}
{Vogt}, S.~S. 1987, \pasp, 99, 1214, \dodoi{10.1086/132107}

\bibitem[{{Vogt} {et~al.}(1994){Vogt}, {Allen}, {Bigelow}, {Bresee}, {Brown}, {Cantrall}, {Conrad}, {Couture}, {Delaney}, {Epps}, {Hilyard}, {Hilyard}, {Horn}, {Jern}, {Kanto}, {Keane}, {Kibrick}, {Lewis}, {Osborne}, {Pardeilhan}, {Pfister}, {Ricketts}, {Robinson}, {Stover}, {Tucker}, {Ward}, \& {Wei}}]{Vogt1994}
{Vogt}, S.~S., {Allen}, S.~L., {Bigelow}, B.~C., {et~al.} 1994, in Society of Photo-Optical Instrumentation Engineers (SPIE) Conference Series, Vol. 2198, Instrumentation in Astronomy VIII, ed. D.~L. {Crawford} \& E.~R. {Craine}, 362, \dodoi{10.1117/12.176725}

\bibitem[{{von Braun} {et~al.}(2011){von Braun}, {Boyajian}, {ten Brummelaar}, {Kane}, {van Belle}, {Ciardi}, {Raymond}, {L{\'o}pez-Morales}, {McAlister}, {Schaefer}, {Ridgway}, {Sturmann}, {Sturmann}, {White}, {Turner}, {Farrington}, \& {Goldfinger}}]{vonBraun2011}
{von Braun}, K., {Boyajian}, T.~S., {ten Brummelaar}, T.~A., {et~al.} 2011, \apj, 740, 49, \dodoi{10.1088/0004-637X/740/1/49}

\bibitem[{{Winn} {et~al.}(2011){Winn}, {Matthews}, {Dawson}, {Fabrycky}, {Holman}, {Kallinger}, {Kuschnig}, {Sasselov}, {Dragomir}, {Guenther}, {Moffat}, {Rowe}, {Rucinski}, \& {Weiss}}]{Winn2011}
{Winn}, J.~N., {Matthews}, J.~M., {Dawson}, R.~I., {et~al.} 2011, \apjl, 737, L18, \dodoi{10.1088/2041-8205/737/1/L18}

\bibitem[{{Wisdom}(2005)}]{Wisdom2005}
{Wisdom}, J. 2005, in AAS/Division of Dynamical Astronomy Meeting, Vol.~36, AAS/Division of Dynamical Astronomy Meeting \#36, 05.08

\bibitem[{{Wright}(2004)}]{wright04}
{Wright}, J.~T. 2004, \aj, 128, 1273, \dodoi{10.1086/423221}

\bibitem[{{Zhao} {et~al.}(2022){Zhao}, {Kunovac}, {Brewer}, {Llama}, {Millholland}, {Hedges}, {Szymkowiak}, {Roettenbacher}, {Cabot}, {Weiss}, \& {Fischer}}]{Zhao2022}
{Zhao}, L.~L., {Kunovac}, V., {Brewer}, J.~M., {et~al.} 2022, Nature Astronomy, \dodoi{10.1038/s41550-022-01837-2}

\bibitem[{{Zhao} {et~al.}(2023){Zhao}, {Kunovac}, {Brewer}, {Llama}, {Millholland}, {Hedges}, {Szymkowiak}, {Roettenbacher}, {Cabot}, {Weiss}, \& {Fischer}}]{zhao23}
---. 2023, Nature Astronomy, 7, 198, \dodoi{10.1038/s41550-022-01837-2}

\end{thebibliography}



\end{document}